\documentclass[11pt,a4paper,english,twoside]{article}

\usepackage{a4wide}
\usepackage{amssymb, amsmath}
\usepackage{graphicx}
\usepackage[all]{xy}
\usepackage{enumerate}
\usepackage{dsfont}
\usepackage{empheq}
\usepackage{cite}
\usepackage{float}
\usepackage{wasysym}
\usepackage{makecell}
\usepackage{colortbl}

\usepackage{microtype}

\usepackage{tocloft}
\usepackage{subcaption}

\usepackage[utf8]{inputenc}
\usepackage[T1]{fontenc}

\usepackage[pdftex,x11names]{xcolor}
\usepackage[pdftex,colorlinks=true,
pdfstartview=FitV,
pdfnewwindow=true,
linktoc = page,
linkcolor= RoyalBlue3,
citecolor= Red3,
urlcolor= RoyalBlue3,
hyperindex=true,
hyperfigures=false]{hyperref}

\newcommand{\beq}{\begin{equation}}
\newcommand{\eeq}{\end{equation}}
\def\bea#1\eea{\begin{align}#1\end{align}}
\def\beal#1\eeal{\begin{subequations}\begin{align}#1\end{align}\end{subequations}}

\newcommand{\eq}[1]{\begin{equation}#1\end{equation}}
\newcommand{\spl}[1]{\begin{split}#1\end{split}}

\newcommand{\boxedeq}[1]{
\begin{equation}
\fbox{
\rule[0.7cm]{0pt}{0pt}
$#1$
\rule[-0.45cm]{0pt}{0pt}
}
\end{equation}
}

\def\d {{\rm d}}

\begin{document}
\numberwithin{equation}{section}

\begin{titlepage}
	\vspace{14pt}
	
	\begin{center}
		
		{\LARGE \bf 
Analytic results for slow-roll curved-space inflation and exponential potentials
		 }\\
		
		\vspace{1.5cm}
		
		{\bf \large Dimitrios Tsimpis${}^{\clubsuit}$ and Govind Venugopal${}^{\diamondsuit}$}
		
		\vspace{0.3cm}

		${}^{\clubsuit}~\!${\it   Institut de Physique des Deux Infinis de Lyon   }\\
		{\it Universit\'e de Lyon, UCBL, UMR 5822, CNRS/IN2P3 }\\
		{\it 4 rue Enrico Fermi, 69622 Villeurbanne Cedex, France  }
		
		\vspace{0.3 cm}

         	${}^{\diamondsuit}~\!${\it  Indian Institute of Science Education and Research (IISER)  }\\
		{\it  Thiruvananthapuram, India  }
		
		\vspace{0.3 cm}

		\texttt{ \href{mailto:tsimpis@ipnl.in2p3.fr}{tsimpis@ipnl.in2p3.fr}, \quad     \href{mailto:vgovind21@iisertvm.ac.in}{vgovind21@iisertvm.ac.in}}
			
		\vspace{1.5 cm}
		
	\end{center}

\vspace{2cm}
\abstract{\noindent We derive analytic templates for the scalar and tensor primordial power spectra  describing 
cosmologies that transition from kinetic dominance to slow-roll inflation  in the presence of spatial curvature.~Our results  
extend recent works  in the literature, allowing us, in particular, to recover the scalar and tensor tilts analytically.~We revisit the case of curvature-assisted single-exponential models in light of this framework. 
In the case of an open universe, the phase space of such models naturally includes cosmologies that start out in a kinetic-dominance regime followed by 
a parametrically controlled quasi-de Sitter phase.  However, they do not fit in the framework of the templates, as their second Hubble slow-roll parameter remains of order one in the quasi-de Sitter regime.
}

\vfill\break
\end{titlepage}
\tableofcontents

\section{Introduction}

Obtaining realistic cosmological models in a well-controlled  top-down construction in string theory is a challenging problem, see \cite{Cicoli:2023opf,Andriot:2026lac} for recent reviews.~The difficulty in embedding de Sitter space in string theory  \cite{Danielsson:2018ztv}, combined with  recent  data compatible with the presence of  a dynamical dark energy
    \cite{DESI:2025zgx, DESI:2025fii}, has led to a  shift of  attention,  within the string-theory community,  to alternative ways to implement accelerated expansion\cite{Agrawal:2018own, Olguin-Trejo:2018zun, Hebecker:2019csg, Cicoli:2020cfj, ValeixoBento:2020ujr, Cicoli:2021fsd, Rudelius:2022gbz, Calderon-Infante:2022nxb, Shiu:2023nph, Shiu:2023fhb, 
Cremonini:2023suw,  Hebecker:2023qke, Freigang:2023ogu, 
VanRiet:2023cca, Shiu:2023yzt, 
Andriot:2024jsh, Shiu:2024sbe, Casas:2024oak, 
 Andriot:2024sif,   Andriot:2025cyi, Dudas:2025ubq}.

 Although it had been known for some time   that asymptotic (i.e.~eternal or semi-eternal) acceleration is possible in eleven-dimensional supergravity \cite{Andersson:2006du}, 
 the link 
  to a four-dimensional effective theory description was not attempted at the time.~Moreover, until recently it was thought that models with transient (i.e.~not asymptotic) acceleration coming from supergravity compactifications could only have a number of e-folds of order one 
  \cite{Townsend:2003fx,Ohta:2003pu,Ohta:2003ie,Ohta:2004wk,Roy:2003nd,Gutperle:2003kc,Emparan:2003gg,Townsend:2003qv, Chen:2003dca,Wohlfarth:2003kw}.

More recently, it was pointed out  in \cite{Marconnet:2022fmx,Marconnet:2025vhj} that asymptotic acceleration  with steep exponentials (i.e.~consistent with the swampland bounds)  is a quite generic feature of different classes of flux compactifications of ten and eleven-dimensional supergravity — the crucial necessary ingredient being an open  universe (i.e. negative spatial curvature).~In addition,  the ``universal cosmology'' models of \cite{Marconnet:2022fmx}  admit   transient acceleration with parametric control of e-folds — a key feature  for the  purposes of the present paper.~Interestingly,  completely  independent arguments have been put forward to argue that an open universe is preferred, or at least naturally realized, in string theory   \cite{Freivogel:2005vv, Bedroya:2025ris}.

In \cite{Andriot:2023wvg} this analysis was revisited from the standpoint of a 4d effective theory, focusing in particular on models with a single-exponential potential.~It was explained therein that the steepness of the exponents typically coming from supergravity (i.e.~in asymptotic regions of string theory, and in accordance with the swampland bounds), together with the negative spatial curvature, guarantee the existence of the attractor fixed point on the boundary of the acceleration region in phase space (denoted $P_1$ in the present paper's phase-space conventions) which is  responsible for the features mentioned in the previous paragraph.

Many recent works have revisited the single-exponential models \cite{Seo:2024fki, Andriot:2025gyr, SanchezGonzalez:2025uco, Bayat:2025xfr, Mosny:2025cyd,Pourtsidou:2025sdd,Arora:2026rho}.~In \cite{Andriot:2024jsh} it was shown that the minimal requirement of radiation followed by matter domination and current accelerated expansion, implies an upper bound   on the exponent (roughly $\sqrt{3}$ for a flat universe, but slightly higher for an open one), and highly restricts the current equation of state parameter of the scalar field.~These results  were recently further extended in \cite{Maki:2026zrn}.~Steep single-exponential models of quintessence appear to be excluded by current observations   \cite{Bhattacharya:2024hep, Alestas:2024gxe,Akrami:2025zlb},  however their multi-field, multi-exponential, spatially-curved extensions offer a much richer phase space structure \cite{Barreiro:1999zs,Li:2005ay,Jarv:2004uk,Liddle:1998jc,Coley:1999mj,Copeland:1999cs,Collinucci:2004iw,Hartong:2006rt,Marconnet:2025vhj,Gallego:2026xqq, Licciardello:2025fhx} and warrant further study. 

Besides quintessence, \cite{Alestas:2024gxe} gave qualitative arguments that curvature-assisted single-exponential potentials  are also in conflict with inflation.~Ref.~\cite{Alestas:2024gxe} only discusses  inflation  in   
the asymptotic fixed-point regime,  where the universe asymptotes the late-time attractor on the boundary of phase space 
(the point $P_1$ mentioned previously) and the acceleration vanishes asymptotically.~However, much more pertinent for inflation are the cosmological histories which correspond to trajectories passing close to the (unstable) curvature-dominated fixed point at the origin of phase space (denoted $P_0$ in the present paper), 
where the universe is in a quasi-de Sitter state.

This is exactly  the context where the parametric control mechanism pointed out  in \cite{Marconnet:2022fmx, Andriot:2023wvg} applies.~More precisely, let $d$ be the minimal distance of a trajectory to the origin of phase space $P_0$, and consider the number of e-folds $\Delta N$ accumulated in the vicinity of $P_0$. 
We can parametrically increase the number of e-folds  accumulated while in quasi-de Sitter expansion, 
by fine-tuning  the initial conditions of the trajectory so that it passes closer to the origin of phase space, 
$$
\boxed{
d\rightarrow0\Rightarrow\Delta N\rightarrow\infty~,~~
\begin{array}{c}
\text{\it while being in a state of quasi-de Sitter expansion}\\
(\text{\it in the vicinity of } P_0)
\end{array}
}$$
By focusing on $P_1$ instead of $P_0$, \cite{Alestas:2024gxe} obtain a type of acceleration which is unsuitable for describing an inflationary phase of the universe.

The present work was motivated by the desire to examine whether the parametric control mechanism described in the previous paragraph can be used to construct viable models of inflation from single-exponential potentials and negative spatial curvature. More specifically, we ask whether such models can be used to produce primordial power spectra (PPS) compatible with data, even at the cost of fine-tuning. A key observation is that the  phase-space of such models 
naturally includes trajectories (cosmologies) that 
start out in a kinetic-dominance (KD) regime followed by a 
parametrically controlled quasi-de Sitter phase.
 This is very similar to the framework of Handley--Thavanesan--Werth \cite{Thavanesan:2020lov}. 

Ref.~\cite{Thavanesan:2020lov} focuses on PPS of cosmological models with a past KD phase followed by a (strict) de Sitter phase,\footnote{In \cite{Thavanesan:2020lov} this phase was called ``ultra-slow-roll'' (USR) inflation. We will refrain from using this terminology in the present paper,  to avoid confusion with other uses of the term USR. We thank Denis  Werth for bringing this to our attention in an earlier version of the paper.} extending the earlier flat-space results of \cite{Contaldi:2003zv} to the case of non-vanishing spatial curvature. At the cost of certain approximations, they were thus able to obtain analytic templates for the scalar PPS of such models.\footnote{We use the term ``template'' 
 to mean an analytic expression for the primordial spectrum, designed to capture the leading physical effects of a class of backgrounds without having to solve the full problem numerically for each model.} Their results were recently extended to analytic templates for the tensor PPS in \cite{Msolla:2025kpd}.

In the present work we extend the results of \cite{Thavanesan:2020lov,Msolla:2025kpd} (see \cite{Gratton:2001gw,Ratra:2017ezv,Bonga:2016iuf,Bonga:2016cje,Akama:2018cqv,Ooba:2017ukj,Kiefer:2021iko,Vigneron:2024bfj,Oikonomou:2026mvp} 
for related works) to the case of cosmologies undergoing a transition from KD to slow-roll (SR). In this way we obtain analytic templates to first order in the SR parameters. In particular, we are able to analytically obtain the tilt of the spectra, which had to be put in ``by hand'' in the approach of \cite{Thavanesan:2020lov,Msolla:2025kpd}. These analytic templates, given in eqs.~\eqref{5.37}, \eqref{7.2} below, are interesting in their own right.

Equipped with this result we revisit the case of single-exponential potentials with negative spatial curvature. We find that these models do not fit in the KD-to-SR framework because, although the first SR parameter can be made arbitrarily small for any number of e-folds, the second SR parameter is of order one. Numerical fits are needed to decide whether these models can be made compatible with observations at the cost of fine-tuning.

The plan of the remainder of the paper is as follows. In Section~\ref{sec:2} we set up the single-scalar curved Friedmann--Lema\^itre--Robertson--Walker (FLRW) system and review the appropriate slow-roll parameters for $K\neq0$. In Section~\ref{sec:srK} we derive the scalar PPS template for a KD-to-SR transition in curved space. In Section~\ref{sec:tensors} we carry out the analogous calculation for tensor modes. In Section~\ref{sec:exp} we apply the framework to curvature-assisted single-exponential potentials and explain why their near-$P_0$ quasi-de Sitter regime lies outside standard SR. We conclude in Section~\ref{sec:conclusions}. Various technical intermediate results can be found in the Appendices.

 \section{Single-scalar model}\label{sec:2}

Our starting point is the 
 4d Einstein action with one minimally-coupled scalar field,
\eq{\label{2sc}
S_{4\d}=\int\d^4 x\sqrt{-g}\left(
\frac{1}{16\pi G}R-\frac12 g^{\mu\nu}  \partial_\mu \varphi \partial_\nu \varphi
  -V(\varphi)
\right)~,}
where $G$ is the 4d Newton's constant.  
We are interested in cosmological solutions of     FLRW  form,
\eq{\label{metric4}
\d s^2 = - \d t^2 + a(t)^2 \left(\frac{\d r^2}{1 -K r^2} + r^2 \d \Omega^2 \right)~;~~~a(t)>0 
~,}
where $a(t)$ is the  (dimensionful) scale factor and $K=+1, -1,0$  correspond to a closed, open, flat  4d universe respectively. 
Assuming homogeneous scalar fields, the   matter equations of motion read, 
\eq{\label{2bgfieldeom}
\ddot{\varphi}
+3H\dot{\varphi}+\partial_{\varphi}V(\varphi)=0  
~,}
where $H:=\frac{\dot{a}}{a}$ is the Hubble parameter, and a dot stands for derivative with respect to the cosmic time $t$.  
The gravitational equations of motion are given by 
  the two  Friedmann equations,\footnote{All  equations of motion are invariant under the rescaling $K\rightarrow \alpha^2 K$, $a\rightarrow \alpha a$, for any constant $\alpha>0$.}
 \eq{\spl{\label{2bgF}
H^2&=\frac{8\pi G}{3}\rho-\frac{K}{a^2}\\
\dot{H}&=-4\pi G  \dot{\varphi}^2+\frac{K}{a^2}
~,}}
where the energy density and pressure are given by, 
\eq{\label{2set}
\rho=\frac12  \dot{\varphi}^2+V(\varphi)~;~~~p=\frac12   \dot{\varphi}^2-V(\varphi)
~.}
Moreover, it can be seen that, together with Eqs.~\eqref{2bgfieldeom}, the first Friedmann equation implies the second one. 
We also define, 
\eq{\label{wdef}
w:=\frac{p}{\rho}=\frac{\frac12   \dot{\varphi}^2-V(\varphi)}{\frac12  \dot{\varphi}^2+V(\varphi)}
~.}
Unless otherwise stated, we will  mostly  work in reduced Planck units, $\hbar=c=8\pi G=1$, cf.~Appendix~\ref{sec:units}.

With these definitions, the following are equivalent conditions for acceleration, 
\eq{\spl{\label{acc}
\ddot{a}>0~,\\
\frac{\d}{\d t}\left(\frac{1}{aH}\right)<0~,\\
V(\varphi)> \dot{\varphi}^2~,\\
\rho+3p<0~,
}}
where the second equation above assumes  $\dot{a}\neq0$, and should be understood as valid away from turning points; 
the third condition can  be shown by combining the  two  Friedmann equations, 
\eq{\label{friedcomb}
\dot{H}+H^2=\frac{\ddot{a}}{a}=-\frac13\left( \dot{\varphi}^2-V(\varphi)\right)
~.}

\subsection{Slow-roll parameters for  $K\neq 0$}

We will now review the appropriate curved-space definitions of SR parameters. These are different from, and less familiar than, their flat-space counterparts. 
All formulas in our treatment of SR in this section are valid for $K\neq0$.

Let us define the first  slow-roll parameter   \cite[Eq.~(69)] {Baumann:2009ds},
\eq{\label{src}
\epsilon:=\frac32(w+1)=\frac{-\dot{H}+\frac{K}{a^2}}{H^2+\frac{K}{a^2}}
~,}
 where for the second equality we used \eqref{2bgF}, \eqref{2set}.~The following conditions are equivalent, 
 assuming positive potential/energy density, 
 cf.~\eqref{wdef}, \eqref{src},
\eq{\spl{\label{srdef}
V(\varphi)\gg\dot{\varphi}^2 ~,\\
w\simeq-1~,\\
\epsilon\simeq 0~.
}}
Moreover, slow-roll corresponds to the spacetime being quasi-de Sitter. Indeed, 
de Sitter space is a solution of the equations of motion \eqref{2bgfieldeom}, \eqref{2bgF},  with scale factor given by,\footnote{In  reduced Planck units,  cf.~Appendix~\ref{sec:units},  the scale factor is dimensionless, and we set,
\eq{\label{3.45}
N:=\ln a~.
}
This amounts to   choosing the origin of the $N$-axis so that $a=1$ at $N=0$; in conventional units this corresponds to $a=l_P$ at $N=0$. } 
\eq{\label{dSsf}
a(t)=\begin{cases}
\frac{1}{\Lambda} \sinh\left[\Lambda (t-t_0)\right]~, &K=-1 \\
 e^{\Lambda (t-t_0)}~, &K=0 \\
\frac{1}{\Lambda}  \cosh\left[\Lambda (t-t_0)\right]~, &K=+1 
\end{cases}}
with  $\varphi=\varphi_0$, $V=3\Lambda^2$,  where $t_0$, $\varphi_0$, $\Lambda$ are real constants, and in the case of open slicing we are assuming $t>t_0$.  Inserting in \eqref{src} the scale factor given in \eqref{dSsf}, we obtain 
$\epsilon=0$, or equivalently $w=-1$. In other words, $\epsilon$ measures the departure from an exact de Sitter space.

A {\it slow-roll expansion} can  be thought of as  an expansion around de Sitter space, with all quantities slowly varying with the number of e-folds. More specifically, $\d Q/\d N$ is considered to be subleading with respect to $Q$, for any quantity $Q(N)$. In other 
words the fractional change of $Q$ is small, 
\eq{\label{3.45b}
\frac{\d}{\d N}\ln|Q|\ll 1
~.}
We also introduce the second slow-roll parameter,
\eq{\label{3.4}
\eta:=\epsilon-\frac{1}{2\epsilon}\frac{\d\epsilon}{\d N}~.
}
Assuming we are in a slow-roll regime, $\epsilon\simeq0$, the smallness of $\eta$ guarantees that $\epsilon$ is quasi-constant over many e-folds, 
i.e. $\d \epsilon/\d N$ is subleading with respect to $\epsilon$, $\eta$. For  exact de Sitter space,  we have $\d\epsilon/\d N=0$ and  $\eta$ is indeterminate.

 Using the  equations of motion,  we obtain the following useful relations, 
 \eq{\label{3.6}
 \epsilon=\frac{\dot{\varphi}^2}{2\left( H^2+\frac{K}{a^2}\right)}~;~~~
 \eta=-\frac{\ddot{\varphi}}{H\dot{\varphi}}
 ~.}

\subsection*{Slow-roll parameters in terms of the potential}

The   potential  and its derivatives can be linked to the SR parameters, 
\eq{\spl{
V&=\Big( H^2+\frac{K}{a^2}\Big)(3-\epsilon)
\\
\frac{1}{H}~\!\partial_\varphi V(\varphi)&=\mp
 \left( H^2+\frac{K}{a^2}\right)^{1/2}\sqrt{2\epsilon}~\!
(3-\eta)~,~~~\text{for}~\pm\dot{\varphi}>0\\
\frac{1}{H^2}~\!\partial^2_\varphi V(\varphi)&=
 \frac{\d{\eta}}{\d N} +(3-\eta)
 \left[
\epsilon+\eta-\frac{K}{(aH)^2}(1-\epsilon)
\right]~.
} }
Alternative  SR  parameters are sometimes defined in terms of the potential and its derivatives,
\eq{\label{sralt}
\epsilon_V:=\frac12 \left(\frac{\partial_\varphi V}{V}\right)^2~;~~~
\eta_V:= \frac{\partial^2_\varphi V}{V}
~.}
These are usually referred to as the {\it potential slow-roll parameters}, to distinguish them from 
the first and second {\it Hubble slow-roll parameters} $\epsilon$, $\eta$ introduced earlier.

Taking \eqref{3.6}-\eqref{sralt} into account we obtain,
\eq{\spl{\label{3.8}
 \left[1+\frac{K}{(aH)^2}\right]
\epsilon_V&=\epsilon \left(\frac{3-\eta}{3-\epsilon}\right)^2
\\
 \left[1+\frac{K}{(aH)^2}\right]
\eta_V&=  \frac{1}{3-\epsilon}~\!\frac{\d{\eta}}{\d N}
 + \frac{3-\eta}{3-\epsilon}
 \left[
\epsilon+\eta-\frac{K}{(aH)^2}(1-\epsilon)
\right]
~.}}
For $K=0$, the  above reduce to the familiar relations  \cite{Baumann:2009ds}, 
 \eq{
 \epsilon_V=\epsilon\left[1+\mathcal{O}(\epsilon,\eta)\right]~;~~~\eta_V=(\epsilon+\eta)\left[1+\mathcal{O}(\epsilon,\eta)\right]~.
 }
However,   for $K\neq0$,  $\eta_V$ is not necessarily small in the slow-roll limit:
\eq{
\eta_V=-\frac{K}{(aH)^2+K}+\mathcal{O}(\epsilon,\eta)~,
}
where we took into account that $\d\eta/\d N$ is subleading with respect to $\epsilon$, $\eta$, cf.~the discussion around  \eqref{3.45b}.

We see that the usual flat-space definition of slow-roll parameters in terms of the potential ceases to be useful in the case of curved space, $K\neq0$.

 \section{Slow-roll inflation and scalar PPS for \boldmath$K\neq0$\unboldmath}\label{sec:srK}

The construction of the scalar template proceeds in three steps:~we first solve the mode equation in the SR regime, then in the preceding KD regime, and finally match the two solutions at the transition conformal time $\tau_c$.~We keep in the main text the ingredients that determine the PPS, while collecting the intermediate background expansions and matching coefficients in Appendix~\ref{app:technical-details}. 
We closely follow the logic and notation of the calculation of the flat-space PPS of \cite{Baumann:2009ds}, which we recall in Appendix \ref{sec:sr1}.

For $K\neq0$, the modified Mukhanov--Sasaki (MS) equation in the form of   \cite{Thavanesan:2020lov} reads, 
\eq{\label{5.7}
v_k''+\left[
\mathcal{K}^2-\left(
\frac{\mathcal{Z}''}{\mathcal{Z}}
+2K+2K\frac{\mathcal{Z}'}{N'\mathcal{Z}}
\right)
\right]v_k=0~,
}
where we define   \cite{Msolla:2025kpd,Hergt:2022fxk},\footnote{No distinction is made in these references between supercurvature ($0<k<1$) and subcurvature modes ($k\geq1$) in the case of an open universe. Although we do not discuss the subtleties of supercurvature modes  \cite{Lyth:1995cw} in the present paper, the plots of Figures \ref{fig:scalar} \& \ref{fig:tensor} only include subcurvature modes.}
\eq{\label{5.Kcal}
\mathcal{K}^2(k):=
\begin{cases}
k^2~, & 0<k\in\mathbb{R}~,\quad K=0,-1~,\\
k(k+2)~, & 3\leq k\in\mathbb{N}~, \quad K=+1~,
\end{cases}
}
with $k$ the comoving wavevector, so that upon Fourier-mode decomposition $-\mathcal{K}^2(k)$ replaces the scalar Laplacian.~Equation \eqref{5.7} was originally written in \cite{Handley:2019anl}  in terms of cosmic time.~The variable $v_k$  is related to 
the gauge-invariant comoving curvature perturbation  via, 
 \eq{ \label{5.11}
 \mathcal{R}_k=\frac{v_k}{\mathcal{Z}}
 ~,}
where,
\eq{\label{5.8}
\mathcal{Z}:=z\left(
\frac{\mathcal{K}^2-3K}{\mathcal{K}^2-3K+K\mathcal{E}}
\right)^{1/2}~;~~~\mathcal{E}:= \frac{\dot{\varphi}^2}{2 H^2 }= \frac{\varphi^{\prime2}}{2 N^{\prime2} } 
~,}
and $z:=a\dot{\varphi}/H$.

The corresponding expansion of the effective mass term in the curved MS equation is given in Appendix~\ref{sec:c.2}. It allows us to rewrite   \eqref{5.7} as follows,
\eq{\label{5.13}
v_k''+\left(
k_\text{sr}^2- \frac{\nu^2-\frac14}{(\tau_0-\tau)^2}\right)
v_k=0~.
}
Here $\tau_0$ is the conformal-time of the asymptotic endpoint of the extrapolated
slow-roll background, at which $a\to\infty$, and 
we have introduced a dynamically-shifted effective wavevector,
\eq{\label{5.14}
 k_\text{sr}^2:= 
 \mathcal{K}^2-\frac13 K\big(8-4\epsilon_0 + 3\eta_0\big) -
 \frac{\epsilon_0}{\mathcal{K}^2 - 3K} K^2
 ~,}
where $\nu:=\frac32+2\epsilon_0-\eta_0$ and  $\epsilon_0:=\left.\epsilon\right|_{N=0}~\!$,  ~\!$\eta_0:=\left.\eta\right|_{N=0}$. 
It can be checked that  $ k_\text{sr}^2>0$ for $K=0,\pm1$ and $|\epsilon_0|,|\eta_0|\ll1$.  
Moreover, the effective wavevector reduces to the one  in   \cite{Thavanesan:2020lov} in the limit $\epsilon_0,\eta_0\rightarrow0$.

Eq.~\eqref{5.13} is valid up to quadratic corrections in the SR parameters and corrections of order $(\tau_0-\tau)^2$, cf.~\eqref{c.11}.
The solution  is given by,
\eq{\label{54.19}
v_k(\tau)= y^{1/2}\big[
c_1H^{(1)}_\nu(y)+c_2H^{(2)}_\nu(y)
\big]
~,
}
where  $H_\nu^{(1)}$ and $H_\nu^{(2)}$ are Hankel functions of
the first and second kind,  $c_1,c_2$ are constants,
and $y:=k_{\rm sr}|\tau_0-\tau|$. 

\subsection{Kinetic dominance}\label{sec:3.1}

In the  KD  regime, $V(\varphi)=0$. The background solution and its expansion are given in Appendix~\ref{sec:c.3}. The result needed here is that the MS equation \eqref{5.7} becomes,
\eq{\label{6.13}
v_k''+\left(
k_\text{kd}^2+ \frac{1}{4\tau^2}\right)
v_k=0~,
}
where  the dynamically-shifted effective  wavevector is given by,
\eq{\label{6.14}
 k_\text{kd}^2:= 
 \mathcal{K}^2-\frac{32}{3} K+
 \frac{24}{\mathcal{K}^2} K^2 ~.}
Eq.~\eqref{6.13} is valid up to corrections of order $\tau^2$, cf.~\eqref{c.14}. 
The solution of \eqref{6.13} is given by, 
\eq{
v_k(\tau)=y^{1/2}\big[
c_1H^{(1)}_0(y)+c_2H^{(2)}_0(y)
\big]
~,
}
where $c_1$, $c_2$ are constants and $y:=k_\text{kd}|\tau|$. 
Imposing the positive-frequency condition \eqref{3.17} for short wavelengths, $k_\text{kd}\tau\gg1$, 
as in \cite{Contaldi:2003zv,  Thavanesan:2020lov}, 
now implies $c_1=0$ and, 
\eq{\label{6.12}
v_k(\tau)=\tfrac12 e^{-\frac{i\pi}{4}} \sqrt{\pi \tau}~\!
H^{(2)}_0(k_\text{kd}\tau)~.
}
It can then be seen that the mode normalization condition \eqref{3.16} is automatically satisfied. 
The validity of this normalization for $K\neq0$ follows
from the quadratic actions of \cite{Handley:2019anl}, as explained at the end of
Appendix~\ref{sec:sr1}.

\subsection{Matching conditions}\label{sec:patching}

The role of the matching calculation is to translate the positive-frequency KD solution into the linear combination of SR modes that determine the asymptotic late-time amplitude. 

First we connect the two regimes (SR and KD), at some conformal transition  time $\tau_c >0$, following   \cite{Contaldi:2003zv, Thavanesan:2020lov}, by imposing 
 continuity of $N(\tau)$ and $N'(\tau)$ at $\tau=\tau_c$. 
 This determines the cosmological background in terms of 
 $\tau_c$, $\Lambda$ and the SR parameters, cf.~Appendix~\ref{sec:c.4}.

 Next,  imposing continuity of $v_k(\tau)$ and $v_k'(\tau)$ at $\tau=\tau_c$, we obtain, 
\eq{\label{5.29mt}
v_k(\tau)=\frac{\sqrt{\pi}}{2} e^{-\frac{i\pi}{4}}\times\begin{cases}
 \sqrt{\tau}~\!
H^{(2)}_0(k_\text{kd}\tau)~,
&\text{for}~0\leq\tau \leq\tau_c\\
\sqrt{\tau_0-\tau}\left[
d_1H^{(1)}_\nu(k_\text{sr}[\tau_0-\tau])+d_2H^{(2)}_\nu(k_\text{sr}[\tau_0-\tau])\right]~,&\text{for}~\tau_c\leq\tau\leq\tau_0
~,\end{cases}
}
with $\tau_0\simeq(3+2\epsilon_0)\tau_c$, cf.~\eqref{5.28}, while the $k$-dependent
coefficients $d_{1,2}$ are given in \eqref{5.30}.

\subsection{The PPS}

Using the asymptotic formula \eqref{4.22} together with \eqref{5.12}, 
\eqref{5.29},  
we obtain the asymptotic late-time amplitude of the comoving curvature perturbation, 
 \eq{\label{1stppseq}
 \mathcal{R}_k^0:=\lim_{\tau\rightarrow\tau_0^-} \mathcal{R}_k\simeq C\left(
1-\frac{2\mathcal{K}^2-7K}{2(\mathcal{K}^2-3K)}\epsilon_0\right)(d_2-d_1) k_{\text{sr}}^{-\nu}
  ~, }
  where
  an  explicit expression of the $k$-dependent factor  $d_2-d_1$ 
can be found in \eqref{5.30.5},  
and, 
  %
 %
  %
 \eq{
 C= \pm\frac{ e^{\frac{i\pi}{4}}\Lambda }{2\sqrt{\epsilon_0}} \left[1+\mathcal{O}(\epsilon)\right]
 ~,}
 is a $k$-independent constant.~We retain the first-order
slow-roll corrections to the $k$-dependence of the spectrum,
but leave the $k$-independent correction to its overall
normalization implicit.~This correction does not affect the
spectral index or the shape of the spectrum normalized at a
pivot scale, and can be absorbed into the fitted scalar
amplitude. Accordingly, 
 we use approximate  
  equalities to mean {\it up to and including the first subleading order in the SR expansion, except for the explicitly
displayed $k$-independent normalization factors}.

 The scalar PPS is given by,
\boxedeq{\label{5.37}
\mathcal{P}_\mathcal{R}=\Delta_\mathcal{R}^2:=\frac{k^3}{2\pi^2}\left|  \mathcal{R}_k^0 \right|^2
=\frac{\Lambda^2}{8\pi^2\epsilon_0}
\left(
1-\frac{2\mathcal{K}^2-7K}{\mathcal{K}^2-3K}\epsilon_0\right)
~\!|d_2-d_1|^2~\!k^{3}k_{\text{sr}}^{-2\nu}
\left[1+\mathcal{O}(\epsilon)\right]
~,}
where the $\left[1+\mathcal{O}(\epsilon)\right]$ factor is $k$-independent. 
The spectrum is therefore parameterized by the free parameters $\Lambda$, $\tau_c$ and the slow-roll parameters $\epsilon_0$, $\eta_0$.

The scalar spectral index is given by,
 \eq{
 n_s=1+\frac{\d \ln\Delta_\mathcal{R}^2}{\d\ln k} 
 ~.}
This  should be calculated for large values of the wavevector $k$, for which, 
\eq{\label{largek}
k_{\text{sr}},k_{\text{kd}}= k \left[1+\mathcal{O}(k^{-1})\right]~;~~~|d_2-d_1| =1+\mathcal{O}(k^{-1})
~,}
 where we took \eqref{5.14}, \eqref{6.14}, \eqref{4.22.5} into account. 
 In this limit we recover the flat-space result,
 \eq{\label{6.26}
 n_s \simeq 1-4\epsilon_0+2\eta_0
 ~.}
The scalar spectrum derived above can be compared against the standard formula of the form, 
\eq{\label{5.42}
\mathcal{P}_\mathcal{R}=A_s\Big(\frac{k}{k_*}\Big)^{n_s-1}
~.}
To that end, we take some representative Planck-2018 base-$\Lambda$CDM values for the scalar amplitude and tilt, evaluated at the usual pivot scale \cite{Planck:2018vyg,Planck:2018jri},  
\eq{\label{planck1}
A_s = 2.1\times 10^{-9}~, ~~n_s=0.965~, ~~\frac{k_*}{a_0}=0.05~\! \text{Mpc}^{-1}~, 
}
with $a_0=0.5\times 10^{5}~\! \text{Mpc}$,  cf.~Appendix~\ref{sec:units}. 
These numbers are used only as reference values; a dedicated $K\Lambda$CDM fit with the curved PPS template is beyond the scope of this paper.

\begin{figure}[H]
\begin{center}
\includegraphics[width=.7\textwidth]{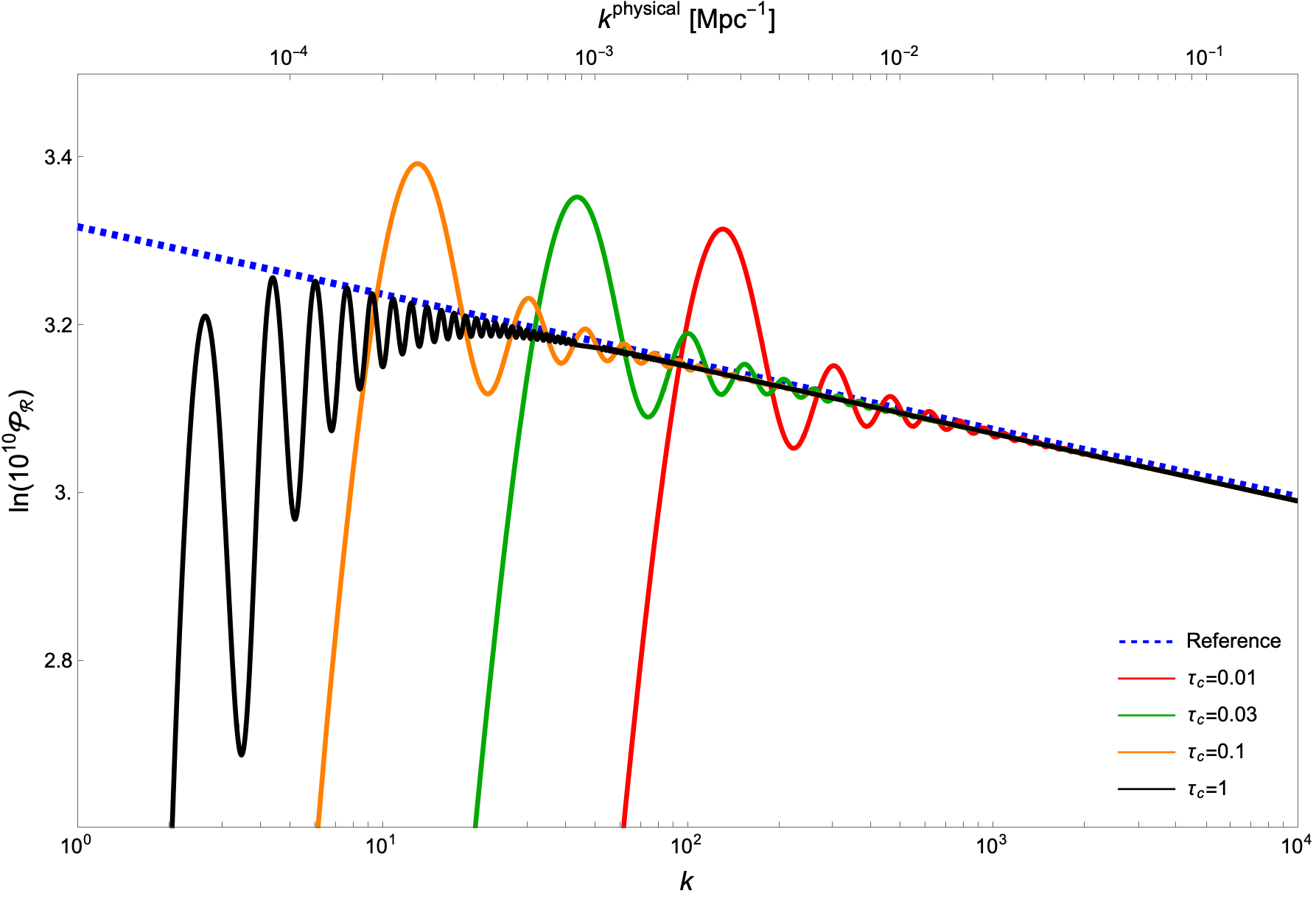}
\caption{Plot of the scalar spectrum \eqref{5.37} for different values of the transition time $\tau_c$. We assume $K=-1$ and the parameter values given in \eqref{planck1comp} of the main text. The 
dashed line is a plot of the standard spectrum \eqref{5.42} for the values given in \eqref{planck1}. The physical wavevector is related to the comoving wavevector $k$ via $k^{\text{physical}}=k/a_0$, where 
we take   $a_0=0.5\times 10^{5}~\! \text{Mpc}$.}\label{fig:scalar}
\end{center}
\end{figure}

Comparing \eqref{5.42} with  \eqref{5.37}  in the $k\gg1$ limit then gives,
\eq{\label{planck1comp}
\Lambda=~2.3\times10^{-5},~~\epsilon_0=0.0024~,~~\eta_0=-0.013~.
}
For these values of the parameters, the plot of the scalar spectrum is depicted in Figure~\ref{fig:scalar}, for different values of the transition time $\tau_c$.




In the usual conformal-time estimate, solving the horizon problem    
requires that the conformal time  elapsed during inflation $\tau_{\rm inf}$ should be greater than the conformal time  afterwards. This requirement 
puts  a bound on the transition time $\tau_c$. 
Indeed, in the instantaneous KD-to-inflation approximation, the conformal time elapsed during inflation is, 
\eq{
 \tau_{\rm inf}\simeq \tau_0-\tau_c~.
}
Using the matching condition \eqref{5.28} we obtain, to leading order in slow-roll,
\eq{
\tau_{\rm inf}\simeq 2\tau_c~.
}
As noted in \cite{Thavanesan:2020lov},  
this automatically exceeds the conformal time elapsed during the  KD  phase, which is $\tau_c$.

On the other hand,  the  condition that the conformal time during inflation should also exceed the conformal time elapsed after inflation,
\eq{\label{horizonconstraintdraft}
 \tau_{\rm after}\lesssim \tau_{\rm inf}\simeq 2\tau_c~, 
}
is  nontrivial.~In imposing this condition, it is important to use the same normalization of conformal time as in our derivation of the spectrum.~In the standard flat-space  cosmological convention, $a_{\rm std}(t_0)=1$, the conformal time 
$ \tau_{\rm after}$  can be roughly taken to be of order $10^4~\!\text{Mpc}$.  
However, our curved-space convention uses a present scale factor $a_0$  equal to the present curvature radius, as explained in Appendix \ref{sec:units}. Thus, 
\eq{
 a(t)=a_0 a_{\rm std}(t)~,~~~
 \d\tau=\frac{\d t}{a(t)}
 =\frac{\d\tau_{\rm std}}{a_0}~.
}
For the value of $a_0$ used in Figure~\ref{fig:scalar}, 
the post-inflationary conformal interval   is of the order 
 $\tau_{\rm after}\simeq 10^4~\!\text{Mpc}/a_0= 0.2$, and  the horizon constraint \eqref{horizonconstraintdraft} translates roughly into $\tau_c\gtrsim0.1$. 
Among the illustrative values shown in Figure~\ref{fig:scalar}, $\tau_c=0.01$ and $\tau_c=0.03$ clearly do not satisfy this estimate.

 \section{Tensor modes for \boldmath$K\neq 0$\unboldmath}\label{sec:tensors}

The curvature-modified   tensor equation in the form of   \cite{Msolla:2025kpd} reads, 
\eq{\label{7.1}
u_k''+\left(
\mathcal{K}^2-
\frac{a''}{a}
+2K
\right)u_k=0~, 
}
where the   variable  $u_k$ is related to the canonically-normalized tensor perturbations via \cite{Baumann:2009ds}, 
 \eq{ \label{9.11}
 {h}_k=2~\!\frac{u_k}{a}
 ~.}
Equation \eqref{7.1} was originally written in \cite{Handley:2019anl}  in terms of cosmic time. 
As in the case of scalar modes, we will solve this equation in the KD and in the SR  regime, and then impose continuity conditions.

\subsection{Kinetic dominance}\label{sec:4.1}

Using the KD background expansion given in Appendix~\ref{sec:c.3}, the curvature-modified tensor equation \eqref{7.1} reduces to,
\eq{\label{7.13}
u_k''+\left(
k_\text{kd,t}^2+ \frac{1}{4\tau^2}\right)
u_k=0~,
}
which is of the same form as \eqref{6.13}, but with a different 
 dynamically-shifted effective wavevector given by,
\eq{\label{7.14}
 k_\text{kd,t}^2:= 
 \mathcal{K}^2+\frac{10}{3} K ~.}
For $K=-1$, this imposes a lower-bound   condition $k>\sqrt{10/3}$ on the wavevectors.

The solution of \eqref{7.13} is given by, 
\eq{
u_k(\tau)=y^{1/2}\big[
c_1H^{(1)}_0(y)+c_2H^{(2)}_0(y)
\big]
~,
}
where $c_1$, $c_2$ are constants and $y:=k_\text{kd,t}\tau$. 
Imposing the positive-frequency condition \eqref{3.17} for short wavelengths, $k_\text{kd,t}\tau\gg1$, 
implies $c_1=0$ and, 
\eq{\label{7.12}
u_k(\tau)=\tfrac12 e^{-\frac{i\pi}{4}} \sqrt{\pi \tau}~\!
H^{(2)}_0(k_\text{kd,t}\tau)~.
}
It can then be seen that the mode normalization condition  \eqref{3.16} is automatically satisfied.

\subsection{Slow-roll inflation}\label{sec:4}

Using the SR background expansion summarized in Appendix~\ref{sec:c.1}, the tensor equation \eqref{7.1} can be written in the same form  as \eqref{5.13}, 
\eq{\label{8.13}
u_k''+\left(
k_\text{sr,t}^2- \frac{\mu^2-\frac14}{(\tau_0-\tau)^2}\right)
u_k=0~,
}
where we have introduced a dynamically-shifted effective wavevector,
\eq{\label{8.14}
 k_\text{sr,t}^2:= 
 \mathcal{K}^2+\frac13 (7+\epsilon_0) K
  ~,}
and $\mu$ is given by, 
\eq{\label{8.15}
\mu:=\frac32+\epsilon_0~.}

The solution is given by,
\eq{\label{8.19}
u_k(\tau)= y^{1/2}\big[
c_1H^{(1)}_\mu(y)+c_2H^{(2)}_\mu(y)
\big]
~,
}
where $c_1$, $c_2$ are constants and   $y:=k_\text{sr,t}|\tau_0-\tau|$. 

\subsection{Matching conditions and PPS}

The tensor matching proceeds exactly as in the scalar case, with the replacements $v_k\to u_k$, $\nu\to\mu$, $k_{\rm kd}\to k_{\rm kd,t}$ and $k_{\rm sr}\to k_{\rm sr,t}$. In particular,
\eq{\label{9.29}
 u_k(\tau)=\frac{\sqrt{\pi}}{2} e^{-\frac{i\pi}{4}}\times\begin{cases}
 \sqrt{\tau}~\!
H^{(2)}_0(k_\text{kd,t}\tau)~,
&\text{for}~0\leq\tau \leq\tau_c\\
\sqrt{\tau_0-\tau}\left[
D_1H^{(1)}_\mu(k_\text{sr,t}[\tau_0-\tau])+D_2H^{(2)}_\mu(k_\text{sr,t}[\tau_0-\tau])\right]~,&\text{for}~\tau_c\leq\tau\leq\tau_0
~,\end{cases}
}
where $D_1$ and $D_2$ are fixed by continuity of $u_k$ and $u_k'$ at $\tau=\tau_c$. Their explicit expressions are given in Appendix~\ref{sec:c.5}.

To calculate the spectrum, we need the late-time asymptotic value of $h_k$. 
Using the asymptotic formula \eqref{4.22} together with \eqref{5.5.6}, \eqref{9.11}, 
\eqref{9.29},  
we obtain, 
 \eq{
 {h}_k^0:=\lim_{\tau\rightarrow\tau_0^-} {h}_k\simeq C (D_2-D_1) k_{\text{sr,t}}^{-\mu}
  ~, }
  where, 
  %
 %
  %
 \eq{
 C= \sqrt{2} ~\!e^{\frac{i\pi}{4}}\Lambda  \left[1+\mathcal{O}(\epsilon)\right]
 ~,}
 is a $k$-independent constant.~As in the scalar case, the $k$-independent first-order correction
to the overall amplitude is left implicit; it does not affect
the tensor spectral index or the shape of the normalized spectrum.

The PPS for the tensor modes is given by,
\boxedeq{\label{7.2}
\mathcal{P}_\mathcal{T}=\Delta^2_\mathcal{T}=2\left(\frac{k^3}{2\pi^2}\right)\left|h^0_k\right|^2
=\frac{2\Lambda^2}{\pi^2}
~\!|D_2-D_1|^2~\!k^{3}k_{\text{sr,t}}^{-2\mu}
\left[1+\mathcal{O}(\epsilon)\right]
~,}
where the $\left[1+\mathcal{O}(\epsilon)\right]$ factor is $k$-independent. 

The  tensor spectral index is given by,
 \eq{
 n_t=\frac{\d \ln\Delta_\mathcal{T}^2}{\d\ln k} 
 ~.}
The tilt  should be calculated for large values of the wavevector $k$,
 for which, 
 \eq{
 |D_2-D_1| =1+\mathcal{O}(k^{-1})
~,}
 where we took  \eqref{4.22.5} into account. 
 In this limit we recover the flat-space result,
 \eq{\label{9.26}
 n_t \simeq -2\epsilon_0
 ~.}
{}Furthermore, from  \eqref{7.2},  \eqref{5.37} we can read off the  tensor-to-scalar  ratio, 
 \eq{\label{9.27}
r=\frac{\Delta^2_\mathcal{T}}{\Delta^2_\mathcal{R}} \simeq  16\epsilon_0
\left(
1+\frac{2\mathcal{K}^2-7K}{\mathcal{K}^2-3K}\epsilon_0\right)\frac{|D_2-D_1|^2}{|d_2-d_1|^2}~\!  \frac{ k_{\text{sr}}^{2\nu} }{k_{\text{sr,t}}^{2\mu}}\left[1+\mathcal{O}(\epsilon)\right]
 ~.}
 The final $k$-independent factor contains the ratio of the
omitted tensor and scalar normalization corrections, which
need not cancel. Thus the overall normalization of $r$ is
specified only at leading order in slow roll.

\begin{figure}[H]
\begin{center}
\includegraphics[width=.7\textwidth]{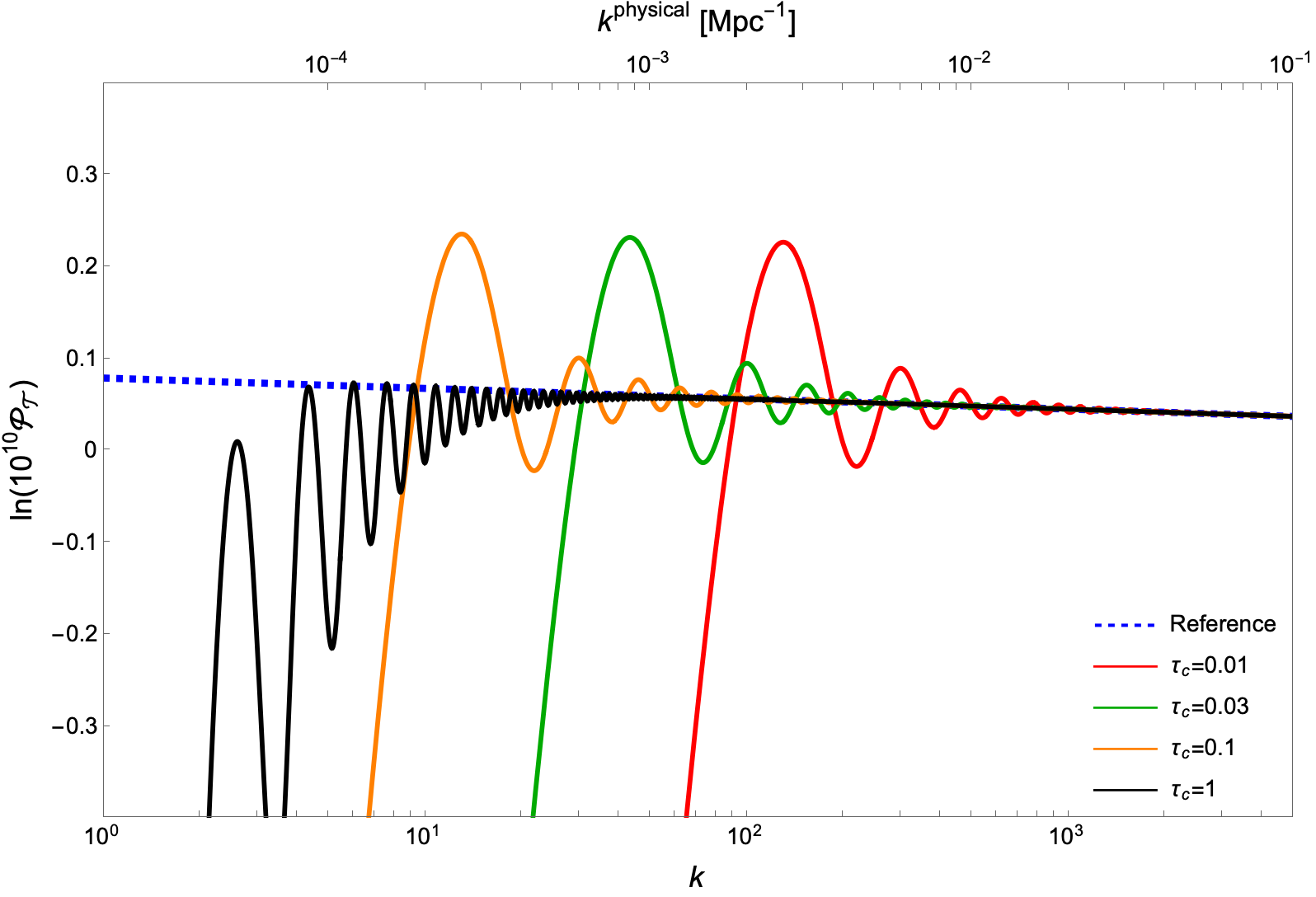}
\caption{Plot of the tensor spectrum \eqref{7.2} for different values of the transition time $\tau_c$. The 
dashed line is a plot of the standard spectrum \eqref{5.42tensor}.~We assume $K=-1$ and the parameter values given in the main text.~The physical wavevector is related to the comoving wavevector via $k^{\text{physical}}=k/a_0$, where 
we take   $a_0=0.5\times 10^{5}~\! \text{Mpc}$.}\label{fig:tensor}
\end{center}
\end{figure}

 The tensor spectrum derived above can be compared against the standard formula of the form, 
\eq{\label{5.42tensor}
\mathcal{P}_\mathcal{T}=A_t\Big(\frac{k}{k_{t*}}\Big)^{n_t}
~.}
To that end we take the following  values from   \cite{Msolla:2025kpd}, 
\eq{\label{planck2}
r:=\frac{A_t }{A_s}=0.05~,  ~~\frac{k_{t*}}{a_0}=0.01~\! \text{Mpc}^{-1}~, 
}
with $\Lambda$, $\epsilon_0$, $\eta_0$ as given in \eqref{planck1comp}. The tensor-to-scalar ratio above 
and the value of $A_s$ given in \eqref{planck1comp} fixes the tensor amplitude, 
\eq{\label{planck3}
A_t = 1.05\times 10^{-10}~. 
}
Moreover, the tilt is determined in  the $k\gg1$ limit by \eqref{9.26}, 
\eq{
n_t=-0.0049
~.}
For these values of the parameters, the plot of the tensor spectrum is depicted in Figure~\ref{fig:tensor}.

\vfill\break




 \section{Exponential potential}\label{sec:exp}
 
 We would now like to revisit the single-exponential models,  in the light of the templates discussed in the first part of the paper.~This is motivated by the fact that  the  phase space of such models 
naturally includes trajectories  that 
start out in a  KD  regime followed by 
a parametrically controlled quasi-de Sitter phase, which might  in principle fit in the KD-to-SR framework of the templates.

\subsection{Dynamical system}

We now recast the single-field background equations of Section~\ref{sec:2} in phase-space form, following the standard dynamical-systems approach to cosmology \cite{we, Bahamonde:2017ize}.  It is useful to introduce the logarithmic slope of the potential \cite{Copeland:1997et},
\eq{\label{26}
\gamma(\varphi):= -\partial_{\varphi}\ln V 
~,}
which would be field-dependent for a generic potential.  For our purposes here we take a single exponential,
\eq{V=V_0e^{-\gamma\varphi}~;~~~V_0\geq0~,}
so that the quantity defined in \eqref{26} is the constant $\gamma$. We choose the sign convention $\gamma>0$, with no loss of generality.  Working with $8\pi G=1$, we use the e-fold number and the normalized kinetic and potential variables\footnote{\label{f1}Negative potentials can be included by defining 
$z=\frac{\sqrt{|V|}}{H\sqrt{3}}$.
With this convention, the terms proportional to $z^2$ in \eqref{system2} and \eqref{1stFried2} acquire the opposite sign. The same sign change in \eqref{210} shows that accelerated expansion is then impossible. Here we restrict our analysis to non-negative potentials.}
\beq
N:= \ln a \ ,\quad x := \frac{\dot{\varphi}}{H \sqrt{6}} \ ,\quad 
z := \frac{\sqrt{V}}{H \sqrt{3}} \ ,\qquad H\neq 0 \ ,\ V\geq0 \ .\label{variables2}
\eeq
The symbol $z$ is used here only as a phase-space coordinate; it should not be confused with the Mukhanov variable $z=a\dot{\varphi}/H$ appearing in Section~\ref{sec:srK} and Appendices~\ref{sec:sr1}, \ref{app:technical-details}.

In these variables the background equations become
\eq{\spl{ \label{system2}
& \frac{\d x}{\d N}=  \sqrt{\frac{3}{2}}\, \gamma \, z^2   + {x} \, \Big( 2{x}^{2} - z^2-2 \Big) \\
&  \frac{\d z}{\d N}= z \left( -\sqrt{\frac{3}{2}}\, \gamma~\!{x}  +  2{x}^{2} -z^2+1 \right)  \ .
}}
The remaining Friedmann equation is encoded in the algebraic condition, 
\eq{
{x}^{2} +z^2 =  1+ \frac{K}{\dot{a}^2} ~.\label{1stFried2}
}
Once imposed on an initial slice, \eqref{1stFried2} is preserved by the flow generated by \eqref{system2}.  Thus \eqref{system2} together with \eqref{1stFried2} is simply the system \eqref{2bgfieldeom}, \eqref{2bgF} written in dimensionless variables.~For a general potential the same rewriting would not close on $(x,z)$ alone, because $\gamma(\varphi)$ would evolve; the single-exponential case is autonomous precisely because $\gamma$ is constant.

Several geometric facts follow directly from the definitions and do not require solving the flow.  Expanding solutions have $z>0$.  In addition, combining \eqref{2bgF} with \eqref{variables2} gives
\eq{\label{210}
\frac{\ddot{a}}{a}=H^2(z^2-2x^{2})
~,}
so the accelerated part of the phase plane is the cone
\eq{\label{52}
\mathcal{A}=\left\{(x,z)\in\mathbb{R}^2~|~z^2>2x^{2}\right\}
~.}
The radial evolution  is governed by
\eq{\label{54}
\frac12\frac{\d }{\d N}  \left({x}^{2} +z^2\right)
=
(z^2-2 {x}^{2} ) (1-x^{2} -z^2)
~.}
In particular, the circle
\eq{\label{ginvs}
\mathcal{C}=\left\{(x,z)\in\mathbb{R}^2~|~x^{2}+z^2=1\right\}
~,}
is invariant under the flow.  A trajectory that starts on this circle remains on it, while trajectories starting inside or outside it remain in the corresponding region.  Through \eqref{1stFried2}, these three possibilities distinguish open, closed and flat FLRW geometries: the interior of $\mathcal{C}$ is the open case, the exterior is the closed case, and the circle itself is the flat case.

Below we focus on the case  $K\leq0$.  Equations \eqref{52} and \eqref{54} then have a simple interpretation: inside the acceleration cone the distance from the origin grows with $N$, whereas outside the cone it decreases.

\subsection{Critical points}\label{sec:critical}

\begin{figure}[H]
\begin{subfigure}{.5\textwidth}
  \centering
  \includegraphics[width=0.7\linewidth]{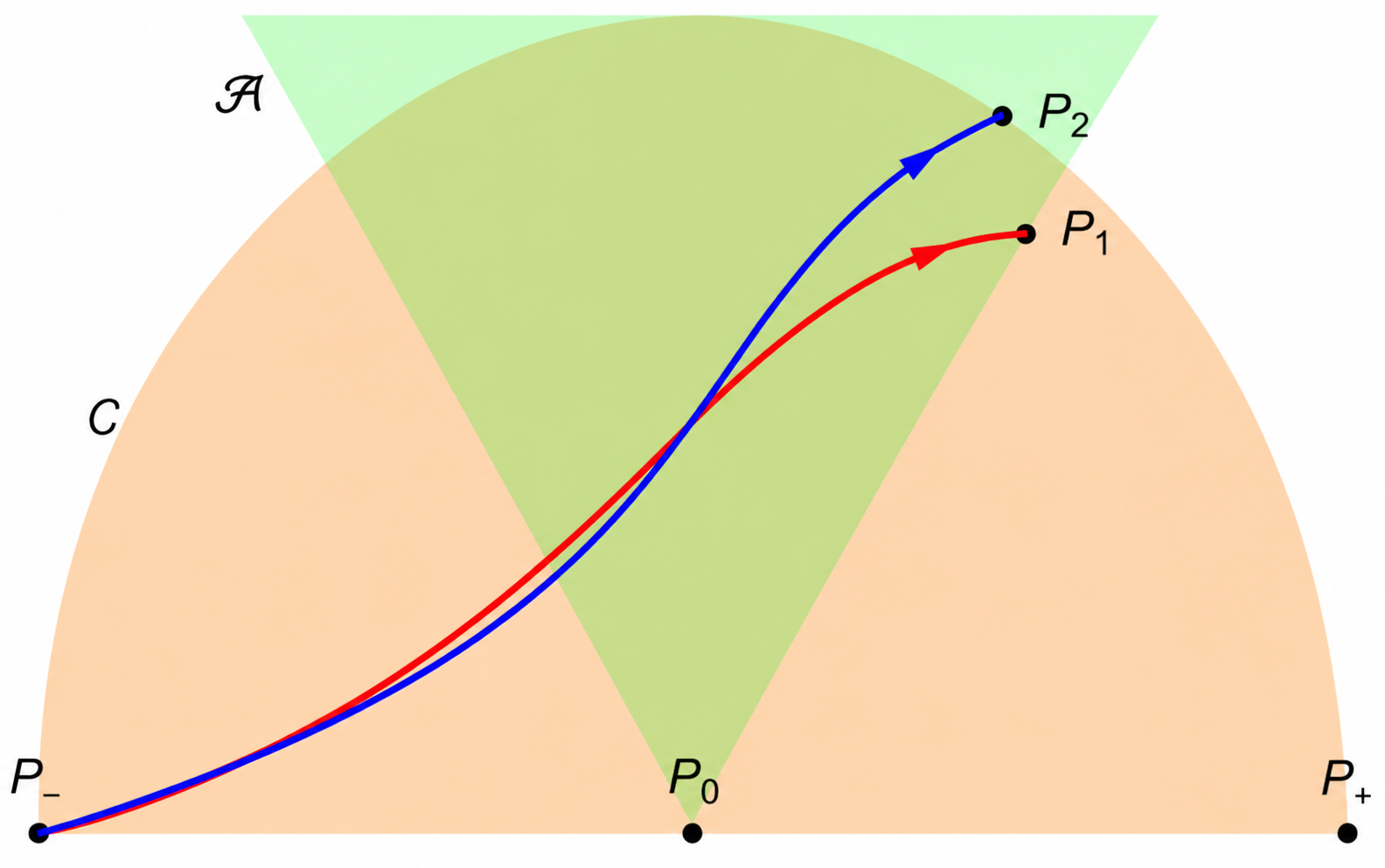}
  \caption{Trajectories asymptoting $P_-$ in the past.}
  \label{fig:2a1}
\end{subfigure}%
\begin{subfigure}{.5\textwidth}
  \centering
  \includegraphics[width=0.69\linewidth]{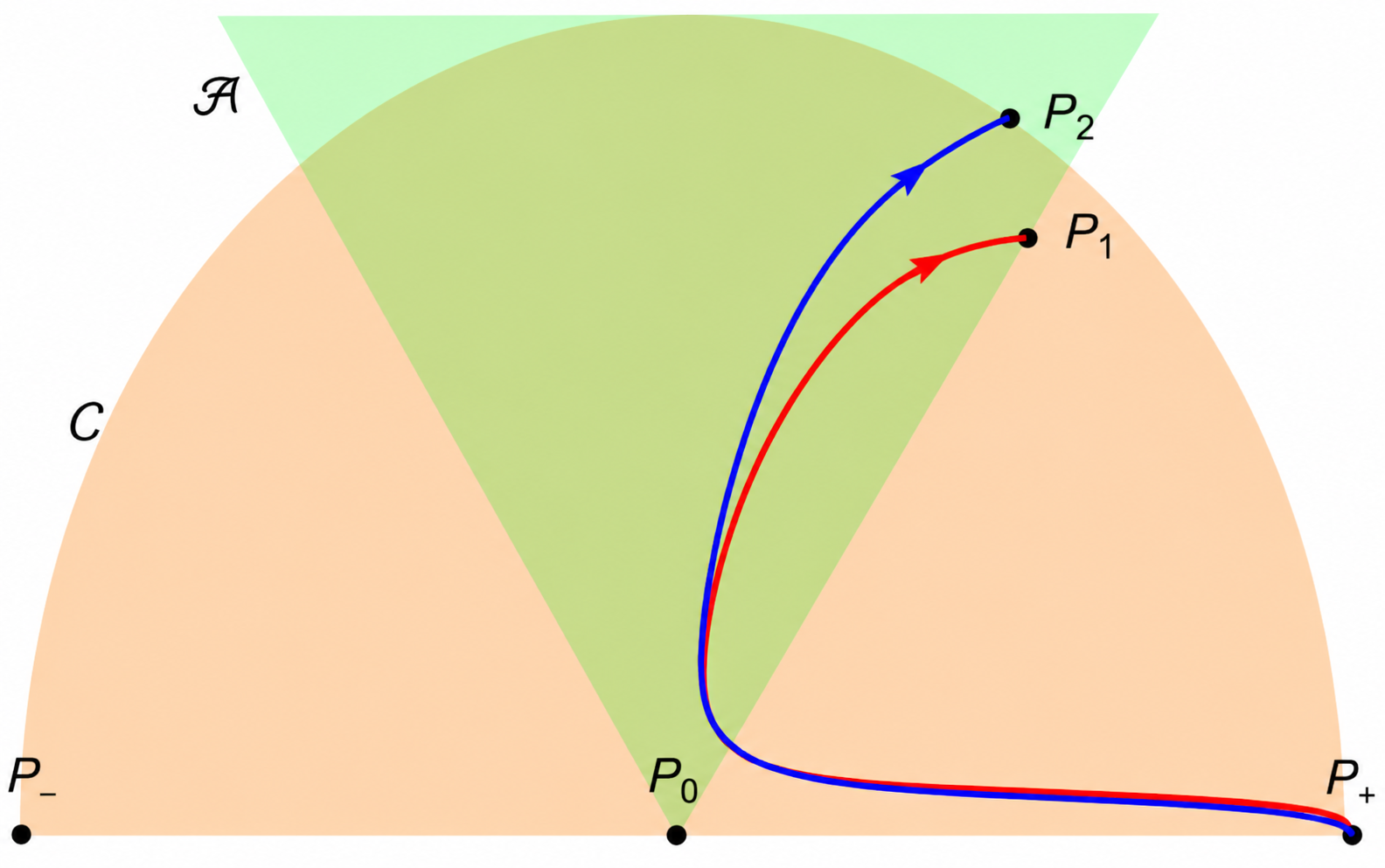}
  \caption{Trajectories asymptoting $P_+$ in the past.}
  \label{fig:3expc}
\end{subfigure}
\caption{Phase portrait of \eqref{system2} subject to \eqref{1stFried2}.  The coordinates of the phase plane are $x$ and $z$.  The orange disk is the open-universe region inside the upper half of $\mathcal{C}$, while the green cone marks the acceleration domain $\mathcal{A}$.  The plot also shows the kinetic-dominated fixed points $P_{\pm}$, the curvature-dominated point $P_0$, the  point $P_2$ with $\gamma^2=1.4$, and the  point $P_1$ on the boundary of the cone with $\gamma^2=2.5$.}
\label{fig:fig_2a1_1et2}
\end{figure}

Solving the fixed-point conditions for \eqref{system2}, with the constraint \eqref{1stFried2}, gives the four cases listed in Table~\ref{tab:fixedpointsfields}, also cf.~Fig.~\ref{fig:fig_2a1_1et2}.

\renewcommand{\arraystretch}{2}
\begin{table}[H]
\begin{center}
\centering
\begin{tabular}{| c | c | c  | c |}
\hline
\cellcolor[gray]{0.9} Point & \cellcolor[gray]{0.9} $(x,z)$ &  \cellcolor[gray]{0.9} Interpretation &    \cellcolor[gray]{0.9} Conditions \\[4pt]
\hline
$P_{\pm}$ & $(\pm1,0)$  & kinetic dominance & $K=0$, $V=0$ \\[7pt]
\hline
$P_0$ &  $(0,0)$  & curvature domination  &  $K=-1$, $V=0$ \\[7pt]
\hline
$P_1$ & $\frac{\sqrt{2}}{\sqrt{3}\gamma}(1,\sqrt{2})$  &  curvature scaling  &  $K=-1$, $\gamma^2>2$ \\[7pt]
\hline
$P_2$ & $\frac{1}{\sqrt{6}}({\gamma},\sqrt{6-\gamma^2})$  &  scalar domination & $K=0$, $\gamma^2<6$ \\[7pt]
\hline
\end{tabular}
\end{center}
\caption {Fixed points of \eqref{system2} compatible with \eqref{1stFried2}.~The table gives their phase-space coordinates, their physical interpretation, and the minimal conditions under which they occur.  For $P_{\pm}$ and $P_0$, the condition $V=0$ may be understood as an asymptotic one; we restrict to  $z\geq0$ and $K\leq0$.}
\label{tab:fixedpointsfields}
\end{table}
\renewcommand{\arraystretch}{1}


%

The nature of these fixed points is made explicit by inserting the corresponding constant values of $(x,z)$ back into the cosmological equations.  The resulting analytic expressions of the scale factors and scalar profiles are collected in Table~\ref{tab:fixedpointssols}.

\renewcommand{\arraystretch}{2}
\begin{table}[H]
\begin{center}
\centering
\begin{tabular}{| c | c | c | c |}
\hline
\cellcolor[gray]{0.9} Point & \cellcolor[gray]{0.9} $a_*(t)$ &  \cellcolor[gray]{0.9} $\varphi_*(t)$ &  \cellcolor[gray]{0.9} Conditions  \\[4pt]
\hline
$P_{\pm} $ & $A_0\,t^\frac{1}{3}$  & $\varphi_{0}\pm\sqrt{\frac23}\ln t$ & $K=0$, $V=0$   \\[2pt]
\hline
$P_0$ &  $t$  & $\varphi_0$ & $K=-1$, $V=0$     \\[2pt]
\hline
$P_1$ & $\frac{~~\gamma}{\sqrt{\gamma^2-2}}\, t$  & $\varphi_{0}+\frac{2}{\gamma}\ln t$  & \makecell{\\[-7pt]$K=-1$ \\[2pt] 
$\varphi_0=\frac{1}{\gamma}\ln  \frac{\gamma^2 V_0}{4}$  \\ [-7pt]{}}   \\
\hline
$P_2$ & $A_0\, t^{\frac{2}{\gamma^2}}$  & ${\varphi}_{0}+\frac{2}{\gamma}\ln t$ & \makecell{\\[-7pt]$K=0$ \\[2pt]
${\varphi}_0=\frac{1}{\gamma}\ln \frac{\gamma^4V_0}{2(6-\gamma^2)}$\\ [-7pt]{}}   \\
\hline
\end{tabular}
\end{center}
\caption {Background solutions obtained at the fixed points.  For $P_{\pm}$ and $P_0$, the vanishing of $V$ is only required asymptoically.~The constants carrying a subscript $0$ are free integration constants except where the table gives them explicitly.  We take $A_0>0$ and use time-translation invariance to place the zero of $a_*(t)$ at $t=0$.}
\label{tab:fixedpointssols}
\end{table}
\renewcommand{\arraystretch}{1}

It is also useful to express the same fixed points in terms of the relative contributions of kinetic energy, potential energy and spatial curvature.  We define
\eq{\label{427}
\rho_{\text{crit}}:=\frac{3H^2}{8\pi G}~;~~~
\Omega_{\text{kin}}:=
\frac{\frac12\dot{{\varphi}}^{~\!2}}{\rho_{\text{crit}}}~;~~~
\Omega_{\text{pot}}:=\frac{V}{\rho_{\text{crit}}}
~;~~~
\Omega_{K}:=-\frac{K}{\dot{a}^2}
~, 
}
so that $\Omega_K>0$ for negative spatial curvature (open universe) in our conventions.  The phase-space variables then give
\eq{\label{428}
\Omega_{\text{kin}}={x}^{~\!2}~;~~~\Omega_{\text{pot}}=z^2~,
}
and the constraint \eqref{1stFried2} becomes
\eq{
\Omega_{\text{kin}}+\Omega_{\text{pot}}+\Omega_{K}=1
~.}
The energy densities at the critical points are shown in Table~\ref{table:energy}.

\vfill\break

\renewcommand{\arraystretch}{2}
\begin{table}[H]
\begin{center}
\centering
\begin{tabular}{| c | c | c | c |}
\hline
\cellcolor[gray]{0.9} Point & \cellcolor[gray]{0.9} $\Omega_K$ &  \cellcolor[gray]{0.9} $\Omega_{\text{kin}}$ &  \cellcolor[gray]{0.9} $\Omega_{\text{pot}}$ \\[4pt]
\hline
$P_{\pm}$ & 0 & 1 & 0 \\[7pt]
\hline
$P_0$ &  1  & 0 & 0 \\[7pt]
\hline
$P_1$ &  $1-\frac{2}{\gamma^2}$    & $ \frac{2}{3\gamma^2}$  & $ \frac{4}{3\gamma^2}$ \\[7pt]
\hline
$P_2$ &  0  & $\frac{\gamma^2}{6}$ & $1-\frac{\gamma^2}{6}$  \\[7pt]
\hline
\end{tabular}
\end{center}
\caption {Energy densities evaluated at the fixed points.  The points $P_{\pm}$ and $P_0$ are limiting regimes of pure kinetic and pure curvature domination, respectively.  At $\gamma=\sqrt{2}$, the entries for $P_1$ coincide with those of $P_2$, and the curvature fraction of $P_1$ goes to zero.}
\label{table:energy}
\end{table}
\renewcommand{\arraystretch}{1}

\begin{figure}[H]
\begin{center}
\includegraphics[width=.7\textwidth]{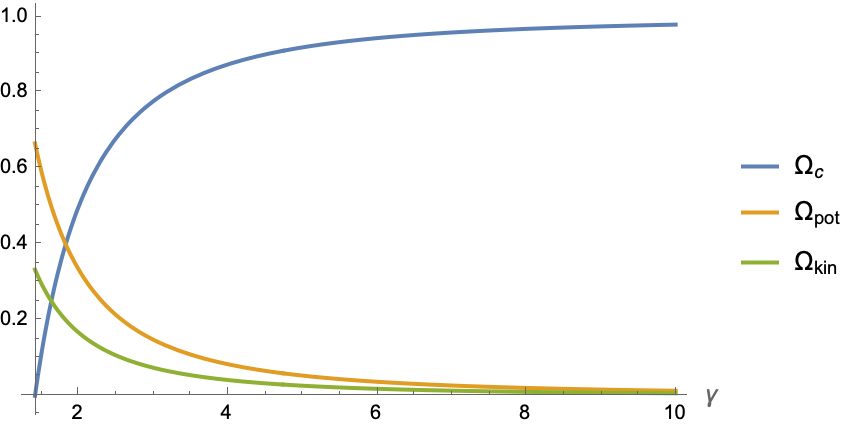}
\caption{Energy fractions at the curvature-scaling point $P_1$ as functions of $\gamma\geq\sqrt{2}$.  The curvature density vanishes in the limit $\gamma\to\sqrt{2}$, while it  becomes dominant for large $\gamma$, while the potential-to-kinetic ratio remains equal to 2.}
\label{fig:engmix}
\end{center}
\end{figure}

For the inflationary question addressed below, $P_1$ is important because it is the late-time attractor when the exponential is steep, $\gamma^2>2$.  This is the range naturally associated with many low-energy string compactification potentials and with the swampland-motivated steepness bounds \cite{Obied:2018sgi,Hebecker:2018vxz,Andriot:2019wrs,Lust:2019zwm,Bedroya:2019snp,Andriot:2020lea,Rudelius:2021oaz,Rudelius:2021azq}.  Figure~\ref{fig:engmix} displays the fixed-point energy fractions as $\gamma$ varies.~Note  that the ratio $\Omega_{\text{pot}}/\Omega_{\text{kin}}$ at $P_1$ is always equal to $2$.




The linear stability results  are summarized in Table~\ref{tab:fixedpointsstab}.  For open trajectories in the interior of $\mathcal{C}$, apart from the special heteroclinic orbit running directly from $P_0$ to $P_1$, the past endpoint is one of the kinetic points $P_\pm$.~The future endpoint is controlled by the steepness of the exponent: $P_1$ is selected when $\gamma^2>2$, whereas $P_2$ is selected when $\gamma^2<2$, cf.~Fig.~\ref{fig:fig_2a1_1et2}.

\renewcommand{\arraystretch}{2}
\begin{table}[H]
\begin{center}
\centering
\begin{tabular}{| c | c | c | c |}
\hline
\cellcolor[gray]{0.9} Point & \cellcolor[gray]{0.9} $(x,z)$ &  \cellcolor[gray]{0.9} Eigenvalues &  \cellcolor[gray]{0.9} Conditions  \\[4pt]
\hline
$P_{\pm} $ & $(\pm1,0)$  & 4,\,$\sqrt{\frac32}(\sqrt{6}\mp\gamma)$ & — \\[7pt]
\hline
$P_0$ &  $(0,0)$  & $-2,\,1$  &  —  \\[7pt]
\hline
$P_1$ & $\frac{\sqrt{2}}{\sqrt{3}\gamma}(1,\sqrt{2})$  & $-1\pm\frac{1}{\gamma}\sqrt{8-3\gamma^2}$  &  $\gamma^2>2$ \\[7pt]
\hline
$P_2$ & $\frac{1}{\sqrt{6}}({\gamma},\sqrt{6-\gamma^2})$  & $\frac12(\gamma^2-6)$, $\gamma^2-2$  & $\gamma^2<6$ \\[7pt]
\hline
\end{tabular}
\end{center}
\caption {Stability analysis for the single-exponential system.~The coordinates and existence conditions are repeated for convenience.  When it exists, $P_1$ is the stable endpoint for $\gamma>\sqrt{2}$; for $\gamma<\sqrt{2}$, the stable endpoint is instead $P_2$.  The fixed-point solutions themselves are given in Table~\ref{tab:fixedpointssols}.}
\label{tab:fixedpointsstab}
\end{table}
\renewcommand{\arraystretch}{1}

The position of $P_1$ on the boundary of $\mathcal{A}$ has an important consequence:~trajectories attracted to $P_1$ must feature an acceleration phase (whether asymptotic or not, depending on their direction of approach) before attaining the boundary.~Furthermore, as emphasized in \cite{Marconnet:2022fmx, Andriot:2023wvg}, 
  the number of e-folds accumulated {\it while being in a state of quasi-de Sitter expansion} increases when the trajectory is chosen to approach the origin $P_0$ more closely.~We will review this mechanism in more detail in Section~\ref{sec:srexp}.

\subsection{Parametric control and slow-roll}\label{sec:srexp}

 In terms of the dynamical system variables we have,
\eq{\label{10.2}
w=\frac{x^2-z^2}{x^2+z^2}~;~~
\epsilon=\frac{3x^2}{x^2+z^2}
~.}
It follows that the cosmology will be in a quasi-de Sitter regime  $\epsilon\lesssim0.01$ as long as the trajectory is inside the cone at the origin with half-angle $\theta\lesssim 3.3^\circ$, cf.~the narrow brown cone in Fig.~\ref{fig:fig_traj}. 
Moreover, from the dynamical system equations and the definition \eqref{3.4} we obtain,
\eq{\label{10.3}
\eta=3\left( 1- \frac{~~\!\gamma z^2}{\sqrt{6}x}\right)
~.}
Let us now examine a   trajectory of interest that asymptotes $P_\pm$ in the past, crosses over into the acceleration cone, and asymptotes $P_1$ in the future, cf.~Figure~\ref{fig:fig_2a1_1et2}.   
In  the vicinity of $P_\pm$, we have, 
\eq{\label{reg1}
\epsilon,~\eta\rightarrow 3~;~~~\frac{1}{aH}\rightarrow0
~,}
as follows from \eqref{10.2}, \eqref{10.3} and $1/(aH)^2=1-x^2-z^2$.

A trajectory approaching the origin  of phase space $P_0$  along the kinetic dominance branch $z\simeq0$ will   have,  
\eq{\label{reg2}
\frac{1}{aH}\simeq1~, ~~\epsilon\simeq 3~,~~\eta\simeq 3~.  
}
On the other hand, consider a
 trajectory which crosses into the acceleration cone at some $N=N_i$, while being  near the origin of phase space: $x(N_i),z(N_i)\simeq0$. 
 For $N\gtrsim N_i$, the trajectory  
can  be  approximated  to any desired accuracy — by fine-tuning  $x(N_i),z(N_i)$ to be closer to the origin — 
by the heteroclinic orbit (the unique trajectory) connecting $P_0$ and $P_1$.~The latter is given by \cite[Eq.~(3.8)]{Marconnet:2025vhj},
 \eq{\spl{\label{xyana}
 \frac{\sqrt{6}}{\gamma}x&=\frac34 {u}^2 -\frac{1}{2^4} (10+3\gamma^2) {u}^4
  +\frac{5}{2^{10}} (112+78\gamma^2+9\gamma^4) {u}^6+ 
\dots
\\
 z&=  {u} -\frac{1}{2^4} (8+3\gamma^2) {u}^3
  +\frac{1}{2^9} (192+160\gamma^2+21\gamma^4) {u}^5+ 
\dots
 ~,}} 
 where  $u:= e^{N-N_0}$, and $N_0$ is a constant. 
 
\begin{figure}[H]
\begin{subfigure}{.5\textwidth}
  \centering
  \includegraphics[width=0.82\linewidth]{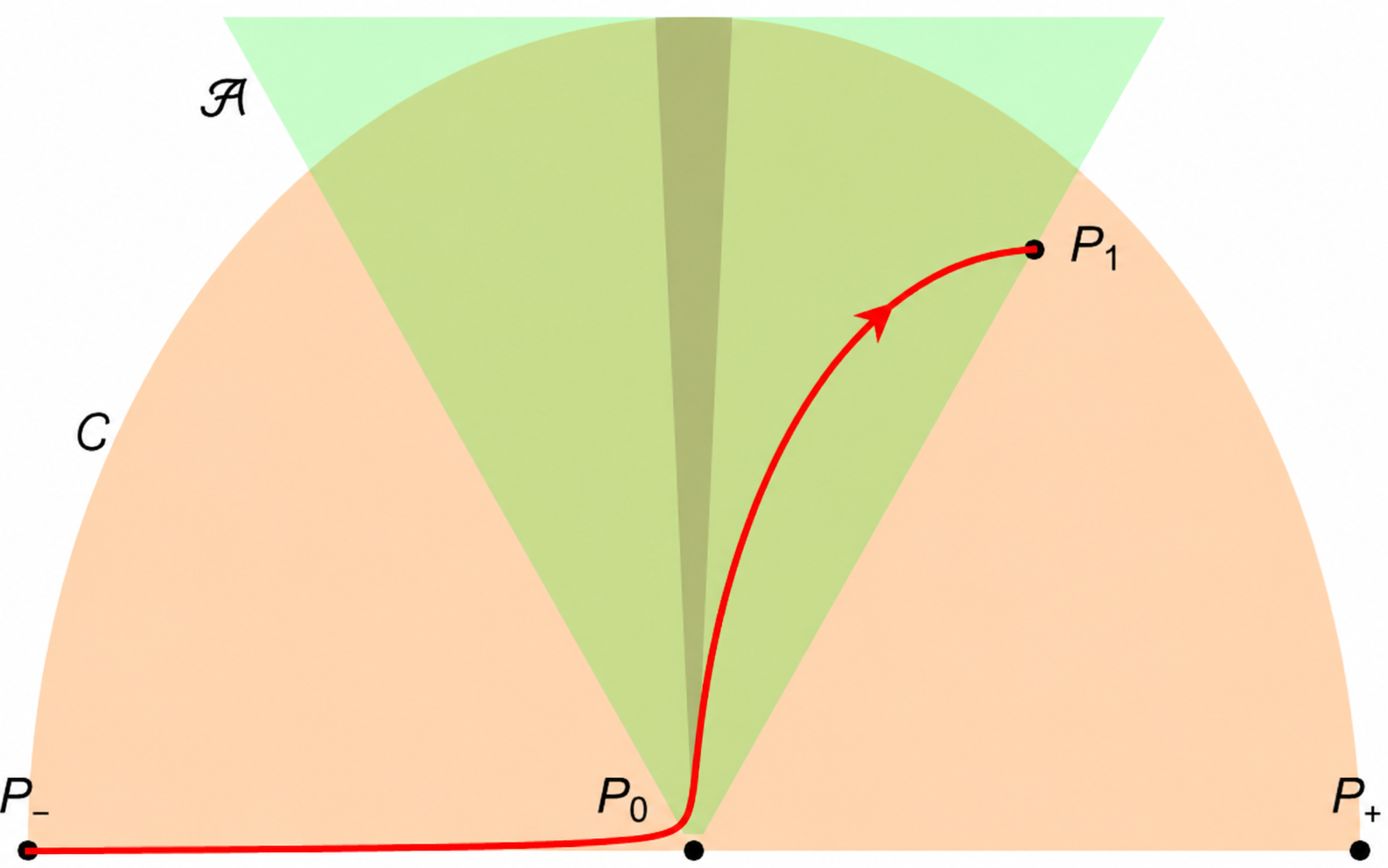}
  \caption{}
  \label{fig:fig_traj_II}
\end{subfigure}%
\begin{subfigure}{.5\textwidth}
  \centering
  \includegraphics[width=1.0\linewidth]{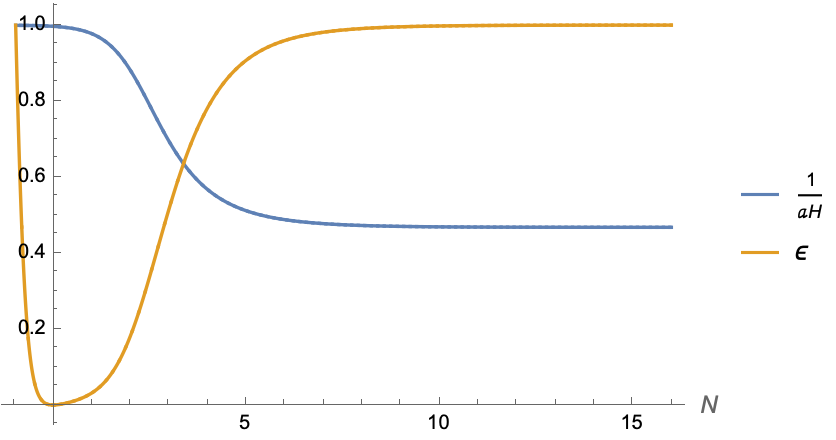}
  \caption{}
  \label{fig:fig_triple_II}
\end{subfigure}
\caption{\!Phase space trajectory asymptoting $P_+$ in the past and $P_1$ in the future (a), and corresponding 
plots of $1/(aH)$, $\epsilon$, as functions of the e-folds parameter $N$ (b). The trajectory is obtained for  an exponential potential with $\gamma=1.6$; it crosses into the acceleration region (the light green cone) at $(x_i,z_i)=(-2.1,3.0)\times 10^{-2}$, which corresponds to $N_i=-0.97$.~The number of e-folds accumulated while being in a quasi-de Sitter state with $\epsilon\lesssim0.01$ (the dark green cone region) is $\Delta N\simeq1$.}
\label{fig:fig1_II}
\end{figure}

\begin{figure}[H]
\begin{subfigure}{.5\textwidth}
  \centering
  \includegraphics[width=0.82\linewidth]{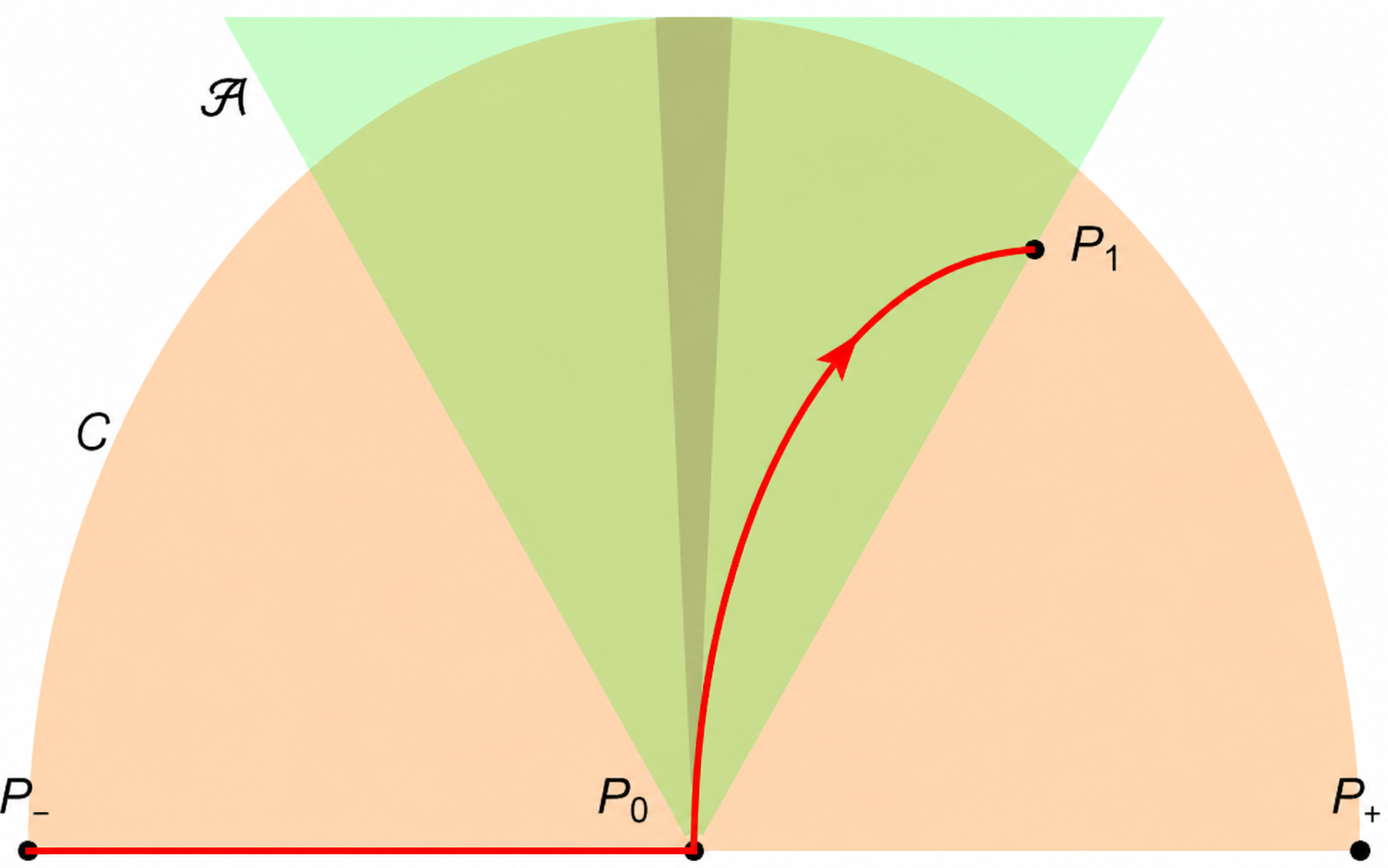}
  \caption{}
  \label{fig:fig_traj_VI}
\end{subfigure}%
\begin{subfigure}{.5\textwidth}
  \centering
  \includegraphics[width=1.0\linewidth]{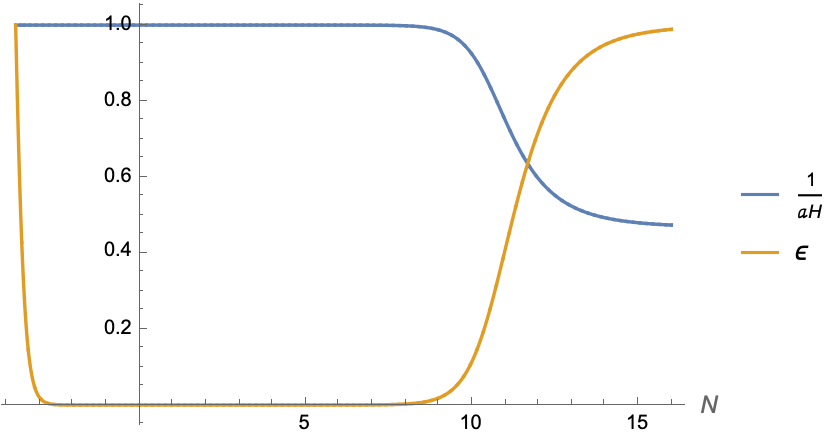}
  \caption{}
  \label{fig:fig_triple_VI}
\end{subfigure}
\caption{\!Phase space trajectory asymptoting $P_+$ in the past and $P_1$ in the future (a), and corresponding 
plots of $1/(aH)$, $\epsilon$, as functions of the e-folds parameter $N$ (b). The trajectory is obtained for  an exponential potential with $\gamma=1.6$; it crosses into the acceleration region (the light green cone) at $(x_i,z_i)=(-3.4,4.9)\times 10^{-7}$, which corresponds to $N_i=-3.7$.~The number of e-folds accumulated while being in a quasi-de Sitter state with $\epsilon\lesssim0.01$ (the dark green cone region) is $\Delta N\simeq12$. The trajectory passes so close to $P_0$, that in the first quadrant it becomes visually indistinguishable from the heteroclinic orbit connecting $P_0$ to $P_1$.}
\label{fig:fig1_VI}
\end{figure}

\begin{figure}[H]
\begin{subfigure}{.5\textwidth}
  \centering
  \includegraphics[width=0.82\linewidth]{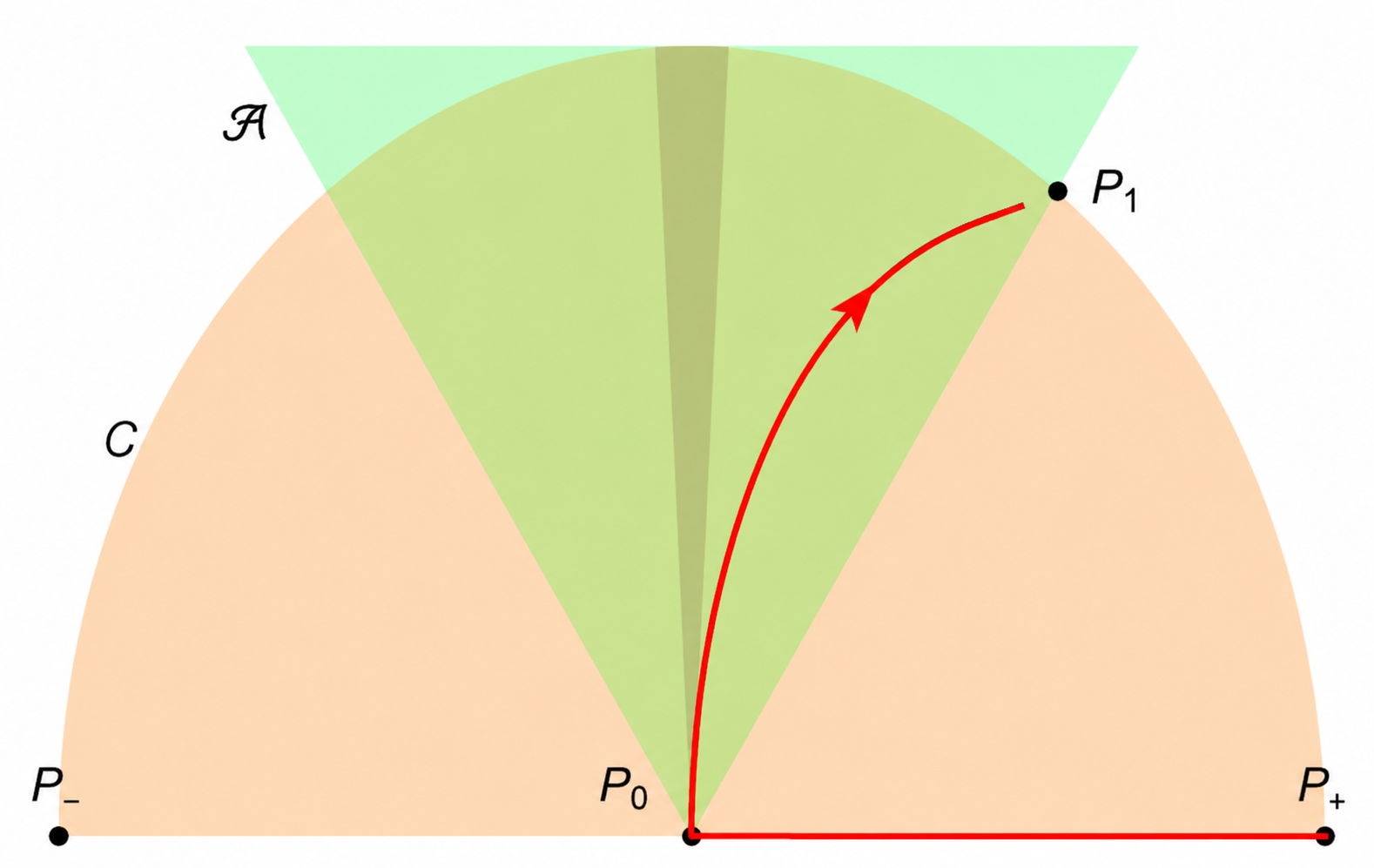}
  \caption{}
  \label{fig:fig_traj}
\end{subfigure}%
\begin{subfigure}{.5\textwidth}
  \centering
  \includegraphics[width=1.0\linewidth]{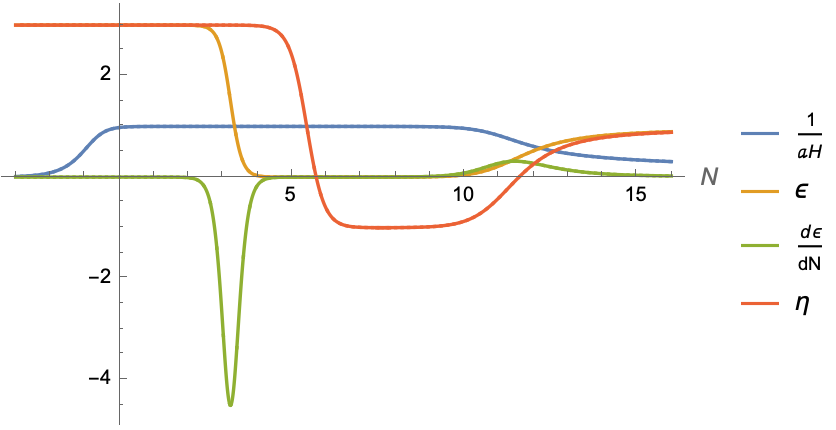}
  \caption{}
  \label{fig:fig_triple}
\end{subfigure}
\caption{\!Phase space trajectory asymptoting $P_+$ in the past and $P_1$ in the future (a), and corresponding 
plots of $1/(aH)$, $\epsilon$, $\eta$ as a function of the e-folds parameter $N$ (b). The trajectory is obtained for  an exponential potential with $\gamma=1.41422$; it crosses into the acceleration region at $(x_i,z_i)=(2.7,3.8)\times 10^{-4}$.~In the asymptotic future we have  $1/(aH)=0.004$.}
\label{fig:fig1}
\end{figure}

 Now consider the number of e-folds $\Delta N=N_f-N_i$, accumulated between $N_i$ and some fixed $N_f$. 
 Let $d:=\min_{(x,z)\in\mathcal{T}}\sqrt{x^2+z^2}$ 
be the minimal distance of the trajectory $\mathcal{T}$ to
the origin of phase space $P_0$. 
Fine-tuning  $d\rightarrow0$, corresponds to taking the limit $x(N_i),z(N_i)\rightarrow0$. 
  In view of \eqref{xyana}, the latter  corresponds to taking the limit  $N_i\rightarrow -\infty$, which in its turn implies $\Delta N\rightarrow\infty$.
 Therefore, as emphasized in \cite{Marconnet:2022fmx, Andriot:2023wvg},  
 we can parametrically control the number of e-folds     accumulated near the origin $P_0$, 
$$
d\rightarrow0\Rightarrow\Delta N\rightarrow\infty~,~~
\text {\it while being in a state of quasi-de Sitter expansion}
$$
This mechanism of parametric control of e-folds is illustrated in Figs.~\ref{fig:fig1_II}, \ref{fig:fig1_VI}: we can increase the number of e-folds $\Delta N$ accumulated while in quasi-de Sitter expansion, by fine-tuning  the initial conditions of the trajectory so that it approaches closer to the origin of phase space.




%
%




%

 From \eqref{xyana} we can also read off, 
 \eq{\label{10.4}
 \epsilon=\frac{9}{32}\gamma^2 u^2+\mathcal{O}(u^4)~;~~~\frac{\d\epsilon}{\d N}=\frac{9}{16}\gamma^2 u^2+\mathcal{O}(u^4)~;~~~
 \eta=-1+\big(\frac23+\frac{1}{2}\gamma^2\big) u^2+\mathcal{O}(u^4)~.
 }
 Therefore, 
in the quasi-de Sitter regime near the origin of phase space, which is  obtained from the above  for $u\simeq0$, we have  
\eq{\label{reg3}
\frac{1}{aH}\simeq1~,~~\epsilon\simeq 0~,~~\frac{\d\epsilon}{\d N}\simeq 0~,~~\eta \simeq -1
~.}
Note that in the quasi-de Sitter regime   both $\epsilon$ and ${\d\epsilon}/{\d N}$ are small. However the latter is of the same order as the former, so that   $|\eta|$  is of order one.

Finally, as the trajectory asymptotes  the vicinity of $P_1$, we have, 
\eq{\label{reg4}
\epsilon,~\eta\rightarrow 1~;~~~\frac{1}{aH}\rightarrow\sqrt{1-\frac{2}{\gamma^2}}~.
}
If we want $1/(aH)\lesssim 0.01$ at late times,  so that the values of the wavevector at horizon crossing span at least two orders of magnitude, 
 we need to fine tune,
\eq{\label{10.1}
\sqrt{2}\leq \gamma \lesssim 1.41428
~.}
Via \eqref{427}, \eqref{reg4},  this would then imply that  $\Omega_K\lesssim 10^{-4}$ near the end of inflation.\footnote{If we consider the vicinity of $P_1$ to roughly correspond to the end of inflation, the value of 
$\Omega_K$ near $P_1$ would have to be several orders of magnitude smaller than $10^{-4}$ for it to subsequently evolve to an observationally acceptable value $\Omega_{K,0}$ today —   for example between 
$10^{-1}$ and $10^{-3}$. 
As can be seen from Table \ref{table:energy}, this  would result in an extreme tuning of $\gamma$ toward $\sqrt2$ from above.
}

The four different regimes \eqref{reg1},  \eqref{reg2},  \eqref{reg3},  \eqref{reg4} can be observed  in  the example of Fig.~\ref{fig:fig1}. In the quasi-de Sitter regime, the parameters 
$\epsilon$ and $\d\epsilon/\d N$ remain $\lesssim0.01$ for $5 \lesssim N \lesssim 9$ in this particular example, however the $\eta$ parameter is close to -1 for the most part.  As explained earlier, 
the number of e-folds for which $\epsilon$ and $\d\epsilon/\d N$ remain small can be made as large as desired, at the cost of fine-tuning the initial data. 
From the phase space point of view, this amounts to choosing a trajectory that passes sufficiently close to the origin $P_0$.




%

In conclusion, this type of single-scalar, single-exponential model does not fit within the  SR framework of Sections~\ref{sec:srK} and~\ref{sec:tensors}, which assumes both $\epsilon\ll1$ and $|\eta|\ll1$.
 It would be interesting to check whether    the exact (numerical) PPS of this type of models   can be made to fit the observational data, at the cost of fine-tuning.

\section{Conclusions}\label{sec:conclusions}

In this work we have derived analytic templates for the scalar and tensor primordial power spectra of curved FLRW cosmologies undergoing a transition from an initial kinetic-dominance phase to a standard slow-roll phase.~Our results extend the KD-to-de Sitter  templates of  \cite{Thavanesan:2020lov} and their tensor analogue  \cite{Msolla:2025kpd} by keeping the first SR  corrections, parameterized by $\epsilon$ and $\eta$, in the inflationary regime. This allows us to recover analytically the scalar and tensor tilts, $n_s$, $n_t$,  rather than inserting them by hand.

To the order considered here, the effect of spatial curvature can again  be captured by dynamical-ly-generated shifts in the  effective wavevectors. 
In the large-$k$ limit the spectra reduce to their standard SR  forms, while at low $k$ they display the expected power suppression  and oscillatory features. 
Moreover, we have shown that 
the requirement that the horizon problem be solved puts a lower bound on the conformal transition time appearing in the templates between KD and SR, roughly $\tau_c\gtrsim 0.1$.

Let us also stress the limitations of the present analytic treatment.~First, our use of a positive-frequency  condition in the KD regime should be understood as an ultraviolet, short-wavelength prescription. This is well justified when the relevant shifted wavevector satisfies $k_{\rm kd}\tau\gg1$, or $k_{\rm kd,t}\tau\gg1$ in the tensor case, so that the mode equation reduces to that of an approximately constant-frequency oscillator and the usual positive-frequency prescription is recovered.~Imposing  the  positive-frequency condition uniquely determines the full classical evolution of all modes.~However,  this  argument justifies imposing it only in the ultraviolet regime; 
%
extending the same prescription to low-$k$ modes is an additional modeling assumption, not a consequence of the high-$k$ limit.

Second,
our analysis is a classical propagation   computation subject to the  initial-condition prescription specified above, using the variables $v$, $u$, of the curved-space Mukhanov--Sasaki formulation, and its tensor analogue, used in \cite{Handley:2019anl}. 
We do not attempt to settle the question of the choice of canonical variables in curved-space quantization, nor the choice of quantum initial conditions to impose for a given variable \cite{Martin:2003sf,Bozza:2003pr,Giovannini:2003it,Agocs:2020yjm,Fulling:1989nb,Fulling:1979ac,Grain:2019vnq,Artigas:2023kyo}.

We then applied this framework to curvature-assisted single-exponential potentials in an open universe.~The  phase space of these models naturally contains trajectories beginning in a KD regime and later entering a region of accelerated expansion. In particular, trajectories passing very close to the curvature-dominated critical point $P_0$ can spend an arbitrarily large number of e-folds in a quasi-de Sitter regime, provided the initial conditions are tuned so that the minimum distance to $P_0$ is sufficiently small. This is the parametric-control mechanism emphasized in \cite{Marconnet:2022fmx,Andriot:2023wvg}.

However, this near-$P_0$ quasi-de Sitter regime does not fall within the standard slow-roll framework developed in Sections~\ref{sec:srK} and~\ref{sec:tensors}: although $\epsilon$ and $\d\epsilon/\d N$ can both be made small near $P_0$, the second slow-roll parameter $\eta$ remains of order one. 
Therefore the analytic KD-to-SR templates derived in this paper cannot be directly applied to these single-exponential trajectories. 

We also note that, without additional ingredients,  these models face a severe flatness constraint: requiring that   the curvature density $\Omega_K$ be sufficiently small near $P_1$ — if we consider this region to roughly correspond to the end of inflation —  so that it subsequently evolves to an observationally acceptable value $\Omega_{K,0}$ today, would result in an extreme tuning of the exponent $\gamma$ toward $\sqrt2$ from above.

  Whether such models can nevertheless yield primordial spectra compatible with observations, at the cost of fine-tuned initial conditions and parameters, remains an open question.~A reliable prediction for their primordial spectra would require solving the exact  mode equations numerically on the corresponding background.~It would also be important to revisit the dependence of the curved primordial spectra on the choice of  initial conditions.

\section*{Acknowledgment} 

D.T.~is grateful to Julien Grain and Denis Werth for valuable discussions.~D.T.~would also like to thank the organizers of the Strings \& Cosmology Meeting at LAPTh Annecy for providing a stimulating environment.~G.V.~would like to acknowledge the hospitality of the Institut de Physique des Deux Infinis de Lyon,  where this work was initiated.~This research is funded in part by the French National Research Agency (ANR) under project no.~ANR-25-CE57-1445-01.

\vfill\break

\appendix

\section{Units}\label{sec:units}

Let us introduce the reduced Planck mass, length and time,
\eq{
 m_P:=\sqrt{\frac{\hbar c}{8\pi G}}~;~~~
 l_P:=\frac{\hbar}{c~\!m_P}=\sqrt{\frac{8\pi G\hbar}{c^3}}~;~~~
 t_P:=\frac{l_P}{c}
~.}
We mostly work in reduced Planck units,
\eq{
\hbar=c=8\pi G=1~,
}
so that $m_P=l_P=t_P=1$.

Contrary to flat universes, for curved universes there is a link between the curvature density and the scale factor.~We will  calibrate, as in \cite{Hergt:2022fxk},  the present-day scale factor $a_0$ from the present-day Hubble parameter $H_0$ and curvature density $\Omega_{K,0}$, 
\eq{\label{a3}
a_0=\frac{c}{H_0}\sqrt{\frac{-K}{\Omega_{K,0}}}
~,}
with $K=-1$ for an open universe. 
We will take $H_0\sim 70$ km~\!s$^{-1}$~\!Mpc$^{-1}$, while $\Omega_{K,0}\sim 10^{-3}$  to $10^{-1}$. This gives a value of $a_0\sim 10^{4}$  to $10^{5}$ Mpc.

 \section{Slow-roll inflation and scalar PPS for \boldmath$K=0$\unboldmath}\label{sec:sr1}

To set the notation, and in order to present the computation in a way that is most readily  generalizable  to the $K\neq0$ case, 
we will here review the derivation of the slow-roll inflation scalar PPS in the flat case 
 $K=0$. We closely follow the conventions of \cite{Baumann:2009ds}. 
 
 From \eqref{src} we obtain, 
\eq{\label{srck0}
\epsilon=- \frac{\d \ln H}{\d N}
~,}
and from \eqref{3.4},
\eq{\label{srck02}
\frac{\d\epsilon}{\d N}=- \frac{\d^2 \ln H}{\d N^2}=2\epsilon(\epsilon-\eta)
~,}
Suppose that the Hubble parameter is equal to $H=\Lambda$ for $N=0$. Moreover, suppose we are in the slow-roll regime, so that $H$ varies slowly as a function of the e-fold number. Taylor-expanding $\ln H$ around $N=0$, taking \eqref{srck0}, \eqref{srck02} into account,  
we obtain,\footnote{Recall that in the SR regime we consider $\epsilon$, $\eta$ to be of the same order, while their variations with respect to $N$ are order $\mathcal{O}(\epsilon^2)$, cf.~the discussion around \eqref{3.45b}.}
 \eq{\label{3.12}
 \ln H(N)=\ln\Lambda-N \epsilon_0+\mathcal{O}(\epsilon^2)\Rightarrow H(N)\simeq\Lambda(1-N \epsilon_0)
 ~.}
 Here and in the following, an approximate  
  equality means {\it up to and including the first subleading order} in the SR expansion. 
Moreover, we set, 
\eq{\epsilon_0:=\left.\epsilon\right|_{N=0}~;~~\eta_0:=\left.\eta\right|_{N=0}
~,}
so that,
\eq{\label{3.15s}
\epsilon=\epsilon_0+2N\epsilon_0(\epsilon_0-\eta_0)+\mathcal{O}(\epsilon_0^3)~.
}
 On the other hand, 
 \eq{\label{3.13}
a H =a \dot{N}=N^\prime~,
 }
 where a prime denotes a derivative with respect to the conformal time $\tau$, defined via $\d t=a\d\tau$. 
 Substituting  \eqref{3.12} into \eqref{3.13} and integrating we thus obtain, 
 \eq{\label{3.14}
  \tau\simeq -\frac{1}{\Lambda}e^{-N}\left[1+(N+1)\epsilon_0\right]~;~~~ N\simeq
  -\ln(\Lambda|\tau|)
    +[1-\ln(\Lambda|\tau|)]\epsilon_0~,}
 where we have dropped an integration constant, which can be recovered by shifting  $\tau$ by a constant. 
 We thus have, 
 \eq{\label{3.18}
 a=e^N\simeq \frac{1}{\Lambda|\tau|}\left(
 1+[1-\ln(\Lambda|\tau|)]\epsilon_0
 \right)~.
 }
 Note that $\tau$ is negative and grows monotonically with $N$;   $\tau\rightarrow-\infty$ corresponds to $N\rightarrow-\infty$, while $\tau\rightarrow 0^-$ corresponds to $N\rightarrow\infty$.

 For $K=0$, the Mukhanov-Sasaki equation reads,
 \eq{\label{3.15}
 v_k^{\prime\prime}+\left(k^2-\frac{z''}{z}\right)v_k=0~,
 }
subject to the normalization  \cite[Eq.~(189)]{Baumann:2009ds},
 \eq{\label{3.16}
 \bar{v}'_k v_k
- \bar{v}_kv_k'=i~,}
and the 
 boundary condition  \cite[Eq.~(192)]{Baumann:2009ds}, 
 \eq{\label{3.17}
 v_k\rightarrow
\frac{1}{\sqrt{2k}}e^{-i k\tau}  ~, }
for $\tau\rightarrow-\infty$, or equivalently, $k|\tau|\gg1$. 
The variable $z$ is defined by,
\eq{\label{4.12}
z:=a\frac{\dot{\varphi}}{H}=\pm a\sqrt{2\epsilon}~,
}
where in the second equality we used \eqref{3.6} with $K=0$. In order to calculate the derivatives of $z$ with respect to $\tau$ we take \eqref{3.14}, \eqref{3.18} into account, together with,
\eq{
\epsilon'=2\epsilon(\epsilon-\eta) N'
~,}
which follows from \eqref{3.4}. We thus obtain,
\eq{\label{3.20}
\frac{z''}{z}\simeq\frac{a''}{a}+(\epsilon-\eta)\Big[
\frac{a''}{a}+\big(\frac{a'}{a}\big)^2
\Big]\simeq\frac{\nu^2-\frac14}{\tau^2}
~,}
where,
\eq{\label{4.15}
\nu:=\frac32+2\epsilon_0-\eta_0~.}
Alternatively we could have simply proceeded from \eqref{3.15s}, \eqref{3.14}, \eqref{3.18} to obtain,
\eq{\label{3.26}
z  \simeq \pm\frac{\sqrt{2\epsilon_0}}{\Lambda|\tau|}
\left[
1+\epsilon_0+(\eta_0-2\epsilon_0)\ln(\Lambda|\tau|)
\right]
~,}
so that,
\eq{\label{4.17}
\frac{z''}{z}\simeq 
\frac{1}{\tau^2}(2+6\epsilon_0-3\eta_0)
~,}
in accordance with \eqref{3.20}. We may also rewrite \eqref{3.26} as,
\eq{\label{3.28}
z  \simeq \pm \sqrt{2\epsilon_0}(1+\epsilon_0)\left(\Lambda|\tau|\right)^{\frac12-\nu}
~.
}
Plugging \eqref{3.20} into \eqref{3.15} implies,
\eq{\label{4.19}
v_k(\tau)=y^{1/2}\big[
c_1H^{(1)}_\nu(y)+c_2H^{(2)}_\nu(y)
\big]
~,
}
where $c_1$, $c_2$ are constants and $y:=k|\tau|$. Furthermore, imposing \eqref{3.17}, 
taking into account the asymptotic formula for the Hankel functions, 
 \eq{\label{4.22.5}
 H_{\nu}^{(1,2)}(y) = \sqrt{\frac{2}{\pi y}}
 ~\! e^{\pm iy}e^{\mp\frac{i\pi}{4}(2\nu+1)}\left[1+\mathcal{O}(y^{-1})\right]
 ~,}
implies $c_2=0$ and, 
\eq{
v_k(\tau)=\tfrac12 e^{\frac{i\pi}{4}(2\nu+1)} \sqrt{\pi|\tau|}~\!
H^{(1)}_\nu(k|\tau|)~.
}
It can then be seen that \eqref{3.16} is automatically satisfied.

 To calculate the spectrum of primordial scalar fluctuations we need the superhorizon limit ($k\ll aH$, or equivalently $\tau\rightarrow0^-$) of 
the  comoving curvature perturbation, 
 \eq{ 
 \mathcal{R}_k=\frac{v_k}{z}
 ~.}
 Using the asymptotic formula for $\text{Re}(\nu)>0$, 
 \eq{\label{4.22}
 H_{\nu}^{(1,2)}(y) = \mp\frac{i\,\Gamma(\nu)}{\pi}\left(\frac{2}{y}\right)^{\nu}\left[1+\mathcal{O}(y^2)\right]
 ~,}
together with \eqref{3.28}, we obtain,
 \eq{
 \mathcal{R}_k^0:=\lim_{\tau\rightarrow0^-} \mathcal{R}_k=C k^{-\nu}
  ~, }
  where the constant, 
 \eq{
 C:=\pm\frac{i\Lambda}{2\sqrt{\epsilon_0}}\left[1+\mathcal{O}(\epsilon)\right]
 ~, }
is $k$-independent. The scalar PPS is given by,\footnote{We follow the conventions for the power spectrum of \cite[Eq.~(145)] {Baumann:2009ds}: $\Delta_\mathcal{R}^2$  as defined here coincides with the dimensionless form of the power spectrum, e.g.~$\mathcal{P}_\mathcal{R}$ of \cite[Eq.~(58)]{Hergt:2022fxk}.
}
\eq{\label{4.25}
\mathcal{P}_\mathcal{R}=\Delta_\mathcal{R}^2:=\frac{k^3}{2\pi^2}\left|  \mathcal{R}_k^0 \right|^2
=\frac{\Lambda^2}{8\pi^2\epsilon_0}~k^{3-2\nu}
\left[1+\mathcal{O}(\epsilon)\right]
~,}
while the scalar spectral index is given by,
 \eq{\label{4.26}
 n_s=1+\frac{\d \ln\Delta_\mathcal{R}^2}{\d\ln k}\simeq 1-4\epsilon_0+2\eta_0
 ~,}
 where we took \eqref{4.15} into account. 
 
 Let us also note that 
the normalization \eqref{3.16} continues to hold  for $K\neq0$, as can be seen from the scalar and tensor quadratic actions of
\cite{Handley:2019anl}. 
Let
$\mathcal{H}:=aH=N'$ and $q_k:=\mathcal{K}^2-3K$.
Expanding the scalar action
\cite[Eq.~(26)]{Handley:2019anl} in real orthonormal spatial
harmonics and using $dt=a\,d\tau$, the action for a real
coefficient $r_k$ of the curvature perturbation becomes, 
\eq{
S_{{\rm s},k}^{(2)}
=\frac12\int d\tau\left[
\mathcal{Z}^2
\left(r_k'-\frac{K}{\mathcal{H}}r_k\right)^2
-z^2q_k r_k^2
\right]~,
\qquad
\mathcal{Z}^2=z^2\frac{q_k}{q_k+K\mathcal{E}}~, 
}
with any spatial degeneracy labels suppressed.~Let us  introduce the real
coordinate $V_k:=\mathcal{Z}r_k$ and define, 
\eq{
A_k:=\frac{\mathcal{Z}'}{\mathcal{Z}}
+\frac{K}{\mathcal{H}}.
}
The integrand then becomes 
$(V_k'-A_kV_k)^2-(q_k+K\mathcal{E})V_k^2$.
Integrating  by parts in order to eliminate the term proportional to $V_kV_k'$ and using the exact
background identity,
\eq{\mathcal{H}'=\mathcal{H}^2(1-\mathcal{E})+K
~,}
we obtain,
up to a boundary term,
\eq{\label{b31}
S_{{\rm s},k}^{(2)}
=\frac12\int d\tau\left[
(V_k')^2-
\left(
\mathcal{K}^2-\frac{\mathcal{Z}''}{\mathcal{Z}}
-2K-\frac{2K}{\mathcal{H}}
\frac{\mathcal{Z}'}{\mathcal{Z}}
\right)V_k^2
\right]~.
}

A similar calculation can be performed for the tensor modes:  starting from
\cite[Eq.~(29)]{Handley:2019anl}, we normalize the real
transverse-traceless harmonics $Q_{k,s}$ by, 
\eq{
\int \text{vol}_3\,
Q^{ij}_{k,s}Q_{ij,k',s'}
=2\delta_{kk'}\delta_{ss'}
~,
}
where $\text{vol}_3$ is the invariant volume element on constant-curvature spatial slices.~The index  $s$  labels the two polarizations, and will be suppressed in the following  
for simplicity. Moreover, we are using a compact notation with Dirac deltas understood for continuous labels.~Taking into account that $-\nabla^2Q_{k}=\mathcal{K}^2Q_{k}$, the 
action for each real tensor mode coefficient $h_{k}$ reads, 
\eq{
S_{{\rm t},k}^{(2)}
=\frac18\int d\tau\,a^2
\left[(h_{k}')^2
-(\mathcal{K}^2+2K)h_{k}^2\right]~.
}
The redefinition $U_{k}:=ah_{k}/2$, followed by integration
by parts, gives, 
\eq{\label{b34}
S_{{\rm t},k}^{(2)}
=\frac12\int d\tau\left[
(U_{k}')^2-
\left(\mathcal{K}^2+2K-\frac{a''}{a}\right)
U_{k}^2
\right]~,
}
again up to a boundary term.  

In the actions \eqref{b31} and \eqref{b34} the canonical momenta are
$\Pi_{V_k}=V_k'$ and $\Pi_{U_k}=U_{k}'$, respectively.
The real oscillator modes should be distinguished from
their complex counterparts $v_k$, $u_k$ used in the main text. 
For example when quantizing the hermitian operator,
\eq{
\widehat V_k
=v_k\widehat a_k+\bar v_k\widehat a_k^\dagger,
\qquad
[\widehat a_k,\widehat a_k^\dagger]=1~, 
}
the canonical commutator
$[\widehat V_k,\widehat\Pi_{V_k}]=i$  gives,
\eq{
v_k\bar v_k'-\bar v_kv_k'=i~,
}
which is precisely \eqref{3.16}. The same argument applies
to the tensor modes $u_k$.

\section{Technical details for the curved templates}\label{app:technical-details}

This appendix collects the intermediate background expansions and matching coefficients used in Sections~\ref{sec:srK} and~\ref{sec:tensors}. They are kept here to streamline the main text.

\subsection{Slow-roll background expansion}\label{sec:c.1}

We here record the first-order SR  background expansion used in Section~\ref{sec:srK}.

For $K\neq0$ in the SR regime, it is not $H$ which is slowly-varying with $N$, but rather the quantity,
\eq{\label{5.1}
\Lambda_{\rm eff}(N):=H\left(1+\frac{K}{(aH)^2}\right)^{1/2}
~.}
Indeed, in the strict de Sitter case, we can see from \eqref{dSsf} that $\Lambda_{\rm eff}$ reduces to the constant $\Lambda$. Moreover, from 
\eqref{src} it follows that,
\eq{\label{5.2}
\epsilon=-\frac{\d\ln \Lambda_{\rm eff}}{\d N}
~.}
Proceeding as in Section \ref{sec:sr1}, we expand,
 \eq{\label{5.3}
 \ln \Lambda_{\rm eff}(N)=\ln\Lambda-N \epsilon_0+\mathcal{O}(\epsilon^2)\Rightarrow (N')^2+K\simeq \Lambda^2 e^{2N}(1-2N \epsilon_0)
 ~,}
where we took \eqref{3.13} into account, and we have set $\Lambda_{\rm eff}(0)=\Lambda$,   $\epsilon_0:=\left.\epsilon\right|_{N=0}$.~Moreover, \eqref{3.4} implies,
\eq{\label{5.6}
\epsilon\simeq\epsilon_0+2N\epsilon_0(\epsilon_0-\eta_0) 
~,}
where $\eta_0:=\left.\eta\right|_{N=0}$.  Recall that  an approximate  
  equality is taken to mean {\it up to and including the first subleading order} in the SR expansion.

Let us  consider the case $K=-1$ first. 
Solving the differential equation \eqref{5.3} order-by-order  in the slow-roll parameter $\epsilon_0$, we obtain, 
 \eq{\label{5.4}
N\simeq
  -\ln(\Lambda\sinh|\tau|)
    +[(\tau+C)\coth\tau-\ln(\Lambda\sinh|\tau|)]\epsilon_0~,}
where we have absorbed an integration constant by a shift in $\tau$.~The second integration constant, $C$, can be determined by requiring that \eqref{5.4} 
agrees with the flat-space result \eqref{3.14} in the $\tau\rightarrow 0$ limit.~This requirement yields $C=0$.~The case $K=+1$ can also be treated similarly, with the result,
 \eq{\label{5.5}
N\simeq \begin{cases}
 -\ln(\Lambda\sinh|\tau_0-\tau|)
    +[(\tau_0-\tau)\coth(\tau_0-\tau)-\ln(\Lambda\sinh|\tau_0-\tau|)]\epsilon_0~,  &\text{if}~K=-1\\
 -\ln(\Lambda\sin|\tau_0-\tau|)
    +[(\tau_0-\tau)\cot(\tau_0-\tau)-\ln(\Lambda\sin|\tau_0-\tau|)]\epsilon_0 ~,     &\text{if}~K=+1~,
\end{cases}
}
where we reinstated an arbitrary integration constant $\tau_0\geq\tau$. Taylor-expanding we get, 
\eq{\label{5.5.5}
N\simeq  -\ln[\Lambda(\tau_0-\tau)]+\frac16 K(\tau_0-\tau)^2+\epsilon_0\left[
1-\ln[\Lambda(\tau_0-\tau)]-\frac16 K(\tau_0-\tau)^2
\right]+\mathcal{O}(\tau_0-\tau)^4
~.}
For $K=0$, we recover the flat-space result \eqref{3.14}. Moreover,
\eq{\label{5.5.6}
a\simeq  
(1+\epsilon_0)
\left(\Lambda|\tau_0-\tau|\right)^{-1-\epsilon_0}\left[1+\mathcal{O}(\tau_0-\tau)^2\right]
~.}

\subsection{Scalar slow-roll effective mass}\label{sec:c.2}

The expansion of the curved Mukhanov--Sasaki variable in the slow-roll regime is as follows.
From \eqref{5.5.5} and the above definitions we obtain to first subleading order in the slow-roll, 
\eq{ \spl{\label{5.12}
\mathcal{Z}&\simeq\pm\frac{\sqrt{2\epsilon_0}}{\Lambda(\tau_0-\tau)}\left(
1+\frac{2\mathcal{K}^2-7K}{2(\mathcal{K}^2-3K)}\epsilon_0+(\eta_0-2\epsilon_0)\ln\left(\Lambda|\tau_0-\tau|\right)
\right)
+\mathcal{O}(\tau_0-\tau)\\
& \simeq \pm \sqrt{2\epsilon_0}
\left(
1+\frac{2\mathcal{K}^2-7K}{2(\mathcal{K}^2-3K)}\epsilon_0\right)
\left(\Lambda|\tau_0-\tau|\right)^{\frac12-\nu}+\mathcal{O}(\tau_0-\tau)
~,}}
 which agrees with the flat-space result \eqref{3.28} for $K=0$. It is also useful to note that,
\eq{\label{5.9}
z =\pm a\sqrt{2\mathcal{E}}~;
~~~
\mathcal{E}=\epsilon\left(
1+\frac{K}{(N')^2}
\right)
~,
}
where we took \eqref{3.6}, \eqref{3.13} into account.

{}Furthermore, 
\eq{\label{c.11}
\frac{\mathcal{Z}''}{\mathcal{Z}}
+2K+2K\frac{\mathcal{Z}'}{N'\mathcal{Z}}
\simeq
\frac{2+6\epsilon_0-3\eta_0}{(\tau_0-\tau)^2}
+\frac13 K\big(8-4\epsilon_0 + 3\eta_0\big) + 
 \frac{\epsilon_0}{\mathcal{K}^2 - 3K} K^2
+\mathcal{O}(\tau_0-\tau)^2
~,}
which agrees with   \cite{Thavanesan:2020lov} in the limit $\epsilon_0,\eta_0\rightarrow0$, and with the flat-space result \eqref{4.17} for $K=0$.

This is the input leading to the slow-roll mode equation in Section~\ref{sec:srK}.

\subsection{Kinetic-dominance background expansion}\label{sec:c.3}

Let us now solve the equations of motion in the kinetic dominance regime, $V(\varphi)=0$. 
In this case, it is straightforward to show that, 
\eq{\spl{\label{6.6}
\varphi&=\varphi_0\pm\sqrt{\frac32}
\begin{cases}
 \ln\tanh\!\left|\tau-\tau_{\text{KD}}\right| ~,& \text{if}~ K=-1\\
 \ln\tan\!\left|\tau-\tau_{\text{KD}}\right| ~,& \text{if}~ K=+1
\end{cases}
\\
N&=N_0+\frac12
\begin{cases}
\ln\sinh\left[2|\tau-\tau_{\text{KD}}|\right]~,& \text{if}~ K=-1\\
 \ln\sin\left[2|\tau-\tau_{\text{KD}}|\right]~,& \text{if}~ K=+1~,
\end{cases}
}}
with  arbitrary constants $N_0$, $\tau_{\text{KD}}$, $\varphi_0$.
We can also give the Taylor expansions around $\tau_{\text{KD}}$,
\eq{\spl{\label{6.5.5}
\varphi&=
\varphi_0\pm\sqrt{\frac32}\ln(\tau-\tau_{\text{KD}})\pm\frac{1}{\sqrt{6}}K(\tau-\tau_{\text{KD}})^2+\mathcal{O}(\tau-\tau_{\text{KD}})^4\\
N&=N_0+\frac12\ln(\tau-\tau_{\text{KD}})-\frac{1}{3}K(\tau-\tau_{\text{KD}})^2+\mathcal{O}(\tau-\tau_{\text{KD}})^4
~.}}
From \eqref{6.5.5} and the   definitions of $\mathcal{E}$, $\mathcal{Z}$, $z$,  cf.~Eqs.~\eqref{5.8}, \eqref{5.9}, 
 we obtain   \cite{Thavanesan:2020lov},  
\eq{\label{c.14}
\frac{\mathcal{Z}''}{\mathcal{Z}}
+2K+2K\frac{\mathcal{Z}'}{N'\mathcal{Z}}
=
-\frac{1}{4(\tau-\tau_{\text{KD}})^2}
+\frac{32}{3} K-
 \frac{24}{\mathcal{K}^2} K^2
+\mathcal{O}(\tau-\tau_{\text{KD}})^2
~.}
This gives the KD mode equation quoted in the main text, upon setting  $\tau_{\text{KD}}=0$, so that $\tau\geq0$.

\subsection{Scalar matching coefficients}\label{sec:c.4}

We will connect the two regimes (SR and KD), at some conformal transition  time $\tau_c >0$, following   \cite{Contaldi:2003zv, Thavanesan:2020lov}. 
In other words,  the cosmological solution is described by \eqref{6.6} (with $\tau_{\text{KD}}=0$ therein) for $0\leq\tau \leq\tau_c$, and by \eqref{5.5} for $\tau_c\leq \tau\leq\tau_0$.

Imposing continuity of $N(\tau)$ and $N'(\tau)$ at $\tau=\tau_c$, to first order in $\epsilon_0$ we obtain, 
\eq{\label{5.28}
\tau_0\simeq\tau_c(3+2\epsilon_0)~;~~~N_0\simeq  \begin{cases} -\ln\left[\Lambda^{1+\epsilon_0}\sinh^{\frac32+\epsilon_0}(2\tau_c)\right]~, &\text{for}~K=-1\\
-\ln\left[\Lambda^{1+\epsilon_0}\sin^{\frac32+\epsilon_0}(2\tau_c)\right]~, &\text{for}~K=+1~.\end{cases}
}
In the limit $\epsilon_0\rightarrow0$, the above is in agreement with  \cite{Thavanesan:2020lov},  if one chooses $\Lambda=\sinh^{-\frac32}(2\tau_c)$, so that $N_0=0$ at zeroth order in $\epsilon_0$. 
Here we keep $\Lambda$ arbitrary.

 Next we impose continuity of $v_k(\tau)$ and $v_k'(\tau)$ at $\tau=\tau_c$, where, cf.~\eqref{54.19}, \eqref{6.12},
\eq{\label{5.29}
v_k(\tau)=\frac{\sqrt{\pi}}{2} e^{-\frac{i\pi}{4}}\times\begin{cases}
 \sqrt{\tau}~\!
H^{(2)}_0(k_\text{kd}\tau)~,
&\text{for}~0\leq\tau \leq\tau_c\\
\sqrt{\tau_0-\tau}\left[
d_1H^{(1)}_\nu(k_\text{sr}[\tau_0-\tau])+d_2H^{(2)}_\nu(k_\text{sr}[\tau_0-\tau])\right]~,&\text{for}~\tau_c\leq\tau\leq\tau_0
~,\end{cases}
}
with $\nu$, $\tau_0$ given in \eqref{4.15}, \eqref{5.28}, and we have introduced new 
rescaled coefficients $d_1$,  $d_2$. We thus obtain, 
\eq{\spl{\label{5.30}
d_1&=\frac{i \pi }{4}\sqrt{(\tau_0-\tau_c)\tau_c}
\Big[
k_{\text{sr}}
H_0^{(2)}(k_{\text{kd}}\tau_c)H^{(2)}_{\nu-1}(k_{\text{sr}}[\tau_0-\tau_c])
-
k_{\text{kd}}
H_1^{(2)}(k_{\text{kd}}\tau_c)H^{(2)}_{\nu}(k_{\text{sr}}[\tau_0-\tau_c])\\
&~~~~~~~~~~~~~~~~~~~~~~~~~~~~~~~~~~~~~~~~~~~~~~~~~~~~+\frac{\tau_0-2\nu\tau_c}{2(\tau_0-\tau_c)\tau_c}
H_0^{(2)}(k_{\text{kd}}\tau_c)H^{(2)}_{\nu}(k_{\text{sr}}[\tau_0-\tau_c])
\Big] 
\\
d_2&=\frac{i \pi }{4}\sqrt{(\tau_0-\tau_c)\tau_c}
\Big[
-k_{\text{sr}}
H_0^{(2)}(k_{\text{kd}}\tau_c)H^{(1)}_{\nu-1}(k_{\text{sr}}[\tau_0-\tau_c])
+
k_{\text{kd}}
H_1^{(2)}(k_{\text{kd}}\tau_c)H^{(1)}_{\nu}(k_{\text{sr}}[\tau_0-\tau_c])\\
&~~~~~~~~~~~~~~~~~~~~~~~~~~~~~~~~~~~~~~~~~~~~~~~~~~~~-\frac{\tau_0-2\nu\tau_c}{2(\tau_0-\tau_c)\tau_c}
H_0^{(2)}(k_{\text{kd}}\tau_c)H^{(1)}_{\nu}(k_{\text{sr}}[\tau_0-\tau_c])
\Big] 
~,}}
so  that,
\eq{\spl{\label{5.30.5}
d_2-d_1&=\frac{i \pi }{2}\sqrt{(\tau_0-\tau_c)\tau_c}
\Big[
-k_{\text{sr}}
H_0^{(2)}(k_{\text{kd}}\tau_c)J_{\nu-1}(k_{\text{sr}}[\tau_0-\tau_c])
+
k_{\text{kd}}
H_1^{(2)}(k_{\text{kd}}\tau_c)J_{\nu}(k_{\text{sr}}[\tau_0-\tau_c])\\
&~~~~~~~~~~~~~~~~~~~~~~~~~~~~~~~~~~~~~~~~~~~~~~~~~~~~-\frac{\tau_0-2\nu\tau_c}{2(\tau_0-\tau_c)\tau_c}
H_0^{(2)}(k_{\text{kd}}\tau_c)J_{\nu}(k_{\text{sr}}[\tau_0-\tau_c])
\Big] 
~.}}
Moreover,  using  \eqref{4.15}, \eqref{5.28}, we have,
\eq{\spl{\label{5.31}
d_1
&=\frac{i \pi\tau_c}{2\sqrt{2}}\left[
k_{\text{sr}}
H_0^{(2)}(k_{\text{kd}}\tau_c)H^{(2)}_{1/2}(2k_{\text{sr}}\tau_c)
-
k_{\text{kd}}
H_1^{(2)}(k_{\text{kd}}\tau_c)H^{(2)}_{3/2}(2k_{\text{sr}}\tau_c)
\right]\left[1+\mathcal{O}(\epsilon)\right]
\\
d_2
&=\frac{i \pi\tau_c}{2\sqrt{2}}\left[
-k_{\text{sr}}
H_0^{(2)}(k_{\text{kd}}\tau_c)H^{(1)}_{1/2}(2k_{\text{sr}}\tau_c)
+
k_{\text{kd}}
H_1^{(2)}(k_{\text{kd}}\tau_c)H^{(1)}_{3/2}(2k_{\text{sr}}\tau_c)
\right]\left[1+\mathcal{O}(\epsilon)\right]
~,}}
which agrees  with  \cite{Thavanesan:2020lov} in the de Sitter  limit $\epsilon\rightarrow0$.

\subsection{Tensor matching coefficients}\label{sec:c.5}

The tensor matching coefficients are obtained from the scalar ones by the replacements
\eq{
 v_k\rightarrow u_k~,~~~\nu\rightarrow\mu~,~~~
 k_{\rm kd}\rightarrow k_{\rm kd,t}~,~~~
 k_{\rm sr}\rightarrow k_{\rm sr,t}~.
}
Together with the tensor mode expansion \eqref{9.29}, these replacements determine $D_1$ and $D_2$ by continuity of $u_k$ and $u_k'$ at $\tau=\tau_c$. We do not repeat the full expressions here, since they are algebraically identical to the scalar coefficients \eqref{5.30} after the substitutions above.

\bibliographystyle{JHEP}

\providecommand{\href}[2]{#2}\begingroup\raggedright\endgroup

\end{document}